\documentclass[12pt]{article}
\usepackage{graphicx}
\usepackage{a4,amsmath,amssymb,cite}
\usepackage{amsfonts}
\begin{document}
\title{ Why Granular Media Are Thermal, \\ and Quite Normal, After All}
\author{Yimin Jiang
\\ Central South University, Changsha 410083, China 
\and Mario Liu
\\ Theoretische Physik, Universit\"{a}t T\"{u}bingen,\\ 72076 T\"{u}bingen, Germany, EC }

\maketitle 

\abstract 
{Two approaches exist to account for granular dynamics: The athermal one takes grains as elementary,  the thermal one considers the total entropy that includes microscopic degrees of freedom 
such as phonons and electrons.  Discrete element method (DEM), granular kinetic theory and athermal statistical mechanics (ASM) belong to the first, granular solid hydrodynamics (GSH) to the second one. 
A discussion of the conceptual differences between both is given here, leading, among others, to the following  insights: \textbullet~While DEM and granular kinetic theory are well justified to take grains as athermal, any entropic consideration is far less likely to succeed. \textbullet~In addition to modeling grains as a gas of dissipative, rigid mass points, it is very helpful take grains as a thermal solid that has been sliced and diced.  \textbullet~General principles that appear invalid in granular media are repaired and restored once the true entropy is included. These abnormalities [such as invalidity of the fluctuation-dissipation theorem, granular temperatures failing to equilibrate, and grains at rest unable to explore the phase space] are consequences of the athermal approximation, not properties of granular media. 
}\\

\noindent
published in: Eur.Phys.J.E (2017) 40: 10;\\ DOI 10.1140/epje/i2017-11497-4\newpage

\tableofcontents

\section{Introduction}

Taking grains as elementary particles interacting via the Newtonian law, 
the discrete element
method (DEM) is the tool of choice for many coming to terms with granular behavior~\cite{CundallStrack,dem1,dem2}. Similarly, granular kinetic theory also assumes that grains are structureless particles undergoing dissipative 
collisions~\cite{kin0,kin1,kin2,kin3,kin4}. 
Both results have consolidated the wide-spread believe of the 
physics community that grains may {\it generally} be approximated as elementary.
Starting from this belief, ``{athermal statistical mechanics}" (ASM) defines a reduced entropy $S_g$ that  does not contain any microscopic degrees of freedom, only those of granular configurations, and assumes it is maximal in equilibrium. One example is the Edward entropy, $S_{Ed}$, given by the number of possibilities grains may be stably packed~\cite{Edw,raphi,1nico,1,2}. 

This is a leap of faith. The irrelevance of microscopics for DEM or the kinetic theory does not imply that the true entropy $S$, of the microscopic degrees of freedom such as phonons and free electrons, is always irrelevant. Brownian motions are negligible because grains are macroscopically large, each containing many microscopic degrees of freedom, such that $S\gg S_g$. Taking $S+S_g$ to be maximal implies equilibrium holds as long as $S$ is maximal,  quite irrespective of the value of $S_g$.

The lack of Brownian motion shows that phonons and electrons do not move grains. But this is insufficient for the conclusion of athermality. Relevant is the question: {\it Which of the two entropies,  $S_g$ or $S$, is being increased by dissipation off equilibrium, during collisions and relaxations, and becomes maximal in equilibrium?} Grains are athermal only if $S_g$ is the answer. However, whenever a system loses energy and heats up measurably -- finger or thermometer, it is $S$ that is being increased.  This is what  grains do in experiments, and if one would trace the lost energy, also in DEM.  Note DEM does not need any entropic considerations, because it already possesses the dissipative terms that push the system toward equilibrium.


ASM takes the reduced entropy $S_g$ to be maximal, and as yet shuns discussion of dissipation. GSH takes $S$ to be maximal in equilibrium, with dissipative terms constructed to increase $S$ off equilibrium. 
Only GSH successfully describes a wide range of granular phenomena, see~\cite{granR1,granR2,granR3,granR4,granRgudehus,granL1,granL2,granL3,exp1.,exp2.,exp3.,exp4.,exp5.,exp6.,exp7.,exp8.,exp9.,PG2013,granRexp}. 
They include fast dense flow, elaso-plastic motion, static stress distribution, propagation of elastic waves, and compaction, And GSH reduces, in appropriate limits,  to the hypoplasticity~\cite{kolymbas1,kolymbas2}, 
Kamrin's nonlocal constitutive relation~\cite{kamrin,kamrin2}, the $\mu(I)$-rheology~\cite{pouliquen1,pouliquen2,pouliquen3}, 
and the kinetic-theory-based hydrodynamic equations of granular gases. 
These results should suffice to convince that the true entropy is a useful quantity, and that GSH presents the appropriate macroscopic framework for understanding the multitude of granular behavior.

GSH is still a qualitative theory, providing a bird's eye view of granular behavior, and approaching a quantitative status only in some select experiments. We are working hard to calibrate the theory's coefficients, making it more realistic.

Its starting postulates -- what we take to be the basic physics underlying granular behavior --  are {\it ``two-stage irreversibility"} and  {\it ``variable transient elasticity''}. The first addresses the three length scales of granular media -- macroscopic, granular and microscopic, and the fact that energy in the macroscopic degrees of freedom first decays into the granular, then the microscopic ones. The second addresses the fact that stresses relax when grains jiggle -- faster the stronger the jiggling is. 

These two concepts  introduce the granular temperature $T_g$ and the elastic strain $u_{ij}$ as state variables. Hereby, $T_g$ quantifies the quickly fluctuating elastic and kinetic energy of the grains, while $u_{ij}$ is static or slowly varying, as it accounts for the grains' coarse-grained elastic deformation. Both $T_g, u_{ij}$ obey equations   accounting for the relaxation toward the equilibrium characterized by maximal true entropy $S$. 
These relaxation equations are therefore what directly link the existence of phonons and free electrons to the flow and stacking of the grains: A flow comes to a halt because there are phonons and electrons to excite, and heat to generate; the stacking is stable because it is the minimum energy state, with all available energy already transfered into heat.
In addition, GSH contains the  conservation equations for mass, momentum and energy, and the balance equation for the true entropy.

In Sec.\ref{2models}, we start by considering a pendulum -- which is, same as a grain, a macroscopic object that may be taken as a rigid mass particle. We ask the question whether it is thermal, using the answer to draw some conclusions. Then we discuss two models for grains: an athermal gas versus a thermal solid that has been sliced and diced, identifying the first with ASM and the second with GSH, and compare both in the context of compaction and tapping, as this is the one subject believed to be well accounted for by ASM. We point to an experiment that agrees with the results of GSH, but would be hard to account for employing ASM. 

In Sec.\ref{sec2}, we first point out that the set of variables of  granular thermodynamics includes the elastic strain $u_{ij}$, the true temperature $T$, and the granular temperature $T_g$. We carefully show how they are defined, and what conceptual pitfalls they entail. Maximizing $S$, we find that granular equilibrium is given by vanishing
$T_g$, uniform  $T$, and the validity of force equilibrium. 

Force equilibrium is especially noteworthy, because it shows that all stable configurations of grains at rest -- those counted by the Edward entropy -- are in equilibrium. They are stable because the true entropy $S$ has a local maximum, while unstable ones in the neighborhood have smaller entropies.  

We also discuss the failure of $T_g$ to equilibrate, showing that given two systems, each with a granular and a true temperature,  the behavior is completely analogous to four systems with four temperatures connected by the same heat currents. There is therefore little unique or incomprehensible about the behavior of $T_g$.  

A conclusion ends this paper, though the appendices also warrant  closer attention. A simplified, minimalist version of GSH is provided there, which streamlines the arguments and expressions, and stresses easy comprehension. It shows why GSH's basic structure is natural, even necessary, and works out its most important ramifications. This renders the paper self-contained, backing up  claims staked in the main text. Therefore, the present paper not only deflects concerns people accustomed to the athermal model have, it also eases their introduction to GSH. 

The simplification of GSH consists of one main point -- taking all transport coefficients to be constant. Generally, they are functions of the density, and constant only if the density is unchanged. If instead the pressure is constant, circumstances are more complicated, because the pressure $P$ is a function of $\rho,u_{ij},T_g$.  If $u_{ij}$ and $T_g$ change with time, the density $\rho$ needs to compensate, rendering the transport coefficients also changing with time. This simplification makes simple, analytical solutions possible, at the price of marring the realism of $P=$ const. experiments. 

After a presentation of GSH, we go on to consider some of its ramifications. We start with compaction, including the reversible and irreversible branch, also the memory effect. Then we examine elasto-plastic motion at given shear rates, especially the approach to the critical state -- a classic experiment in soil mechanics. Next, we discuss shear jamming -- the fact that a granular system becomes jammed upon shearing~\cite{SJ1,SJ2,SJ3} -- showing that it is described by the same general solution as the approach to the critical state, albeit with altered initial conditions. 

Elasto-plastic motion at given shear stresses is equally interesting. The phenomena accounted for include the difference between the angle of repose and stability,  shear bands, and the observation of a divergent shear strain. To account for the latter, Nguyen et al.~\cite{aging} 
borrowed concepts such as {\em fluidity, aging} and {\em rejuvenation parameter} from the glassy dynamics, although these are undefined, at most vague notions  in granular media. As will be shown, they are,  respectively, $T_g$,  its relaxation rate, and its production rate, and the glassy dynamics turns out to be a reduced, scalar version of  GSH in the limit of constant stresses. 

Finally, considering fast dense flow, we show that both the $\mu(I)$-rheology and Kamrin's nonlocal constitutive relation are natural consequences of GSH. All these demonstrate the ease and usefulness of a comprehensive, thermal  approach.

A word on the {\it hydrodynamic formalism}, with the help of which GSH was derived. Developed by Landau in the context of superfluidity~\cite{Khal,LL6} and 
introduced to complex fluids by de Gennes~\cite{deGennes}, it holds a special place in  modeling, for which the dichotomy between realism and comprehension exists. Constitutive relations typically focus on the first, while hydrodynamic theories usually enable the latter. This is because hydrodynamic theories are set up starting from comparatively few initial inputs, all derived from the understanding of the system's  basic physics. (Examples are  the quantities being conserved and continuous symmetries being spontaneously broken. In the case of GSH, as explained, 
it is {\it two-stage irreversibility} and  {\it variable transient elasticity}.)
One then employs general principles -- conservation laws, Galilean invariance, and the second law of thermodynamics -- to confine the structure of the theory. As a result, a hydrodynamic theory leaves little leeway (typically a handful of scalar functions) to fit the wide range of experiments of a given system. When constructing a hydrodynamic theory,  it is therefore quickly obvious if the initial inputs are wrong -- as is frequently the case. If not, the theory develops predictive power and the capability to become realistic. 
In comparison, constitutive models are constructed relying on the accumulated data from specific experiments. 
It is highly realistic, though less predictive in untested circumstances. 
\\

\noindent
{\bf Notations:}\\
We take $\partial_t a\equiv\frac\partial{\partial t}a$ for any $a$, the velocity as $v_i$,
\\and denote the strain rate as  $v_{ij}\equiv\frac12(\nabla_iv_j+\nabla_jv_i)$, \\
its traceless part as $v^*_{ij}$, with
$v_s\equiv\sqrt{v^*_{ij}v^*_{ij}}\,\,\, (\equiv||v^*_{ij}||$).\\
In addition, we  denote the elastic strain $\varepsilon^{ela}_{ij}$ as $u_{ij}$, 
\\ the elastic stress as $\pi_{ij}$, the Cauchy or total stress as $\sigma_{ij}$. \\
Finally, with $u^*_{ij}, \pi^*_{ij}, \sigma^*_{ij}$ traceless, we take
\\
$u_s\equiv\sqrt{ u^*_{ij}u^*_{ij}}$, $\quad\pi_s\equiv\sqrt{ \pi^*_{ij}\pi^*_{ij}}$, $\quad\sigma_s\equiv\sqrt{ \sigma^*_{ij}\sigma^*_{ij}}$,\\
and $\Delta\equiv -u_{\ell\ell}$, $P_\Delta\equiv\pi_{\ell\ell}/3$, $P\equiv\sigma_{\ell\ell}/3$.

\section{Two Opposite Models for Grains\label{2models}}
There are two pictures that we associate with grains, a gas of elementary but macroscopic particles, and a block of rock that has been sliced and diced. The first is athermal, the second thermal.  Both 
work well within their respective range of validity.  The simple example of a pendulum is helpful to find out what they are. 
\subsection{Is a Pendulum athermal?}
A pendulum is, like a grain, a macroscopic object. Its linearized equation of motion, appropriate for small amplitudes, reads (with $\theta$ the pendulum angle, $g$ the gravitational constant, $l$ the length of the string, and $\alpha$ the friction coefficient)
\begin{equation}
\ddot\theta+\alpha \dot\theta+\theta g/l=0.
\end{equation}
Given this equation,  one can calculate the pendulum's motion, its return to equilibrium hanging down, with no need to ever consider its entropy $S$. 

Nevertheless, a pendulum is not athermal, because the frictional force $\alpha \dot\theta$ increases $S$, of the pendulum itself and the surrounding air. 
In fact, one can derive this force by starting from the second law of thermodynamics, that $S$ can only increase, until it is maximal in equilibrium. The total and conserved energy $E$ is given by the potential, kinetic and heat contributions,  
$E=\frac12Mgl\theta^2+\frac12Ml^2\dot\theta^2+\int T{\rm d}S$, or
$\dot E=0=Mgl\theta\dot\theta+Ml^2\dot\theta\ddot\theta+T\dot S$. Inserting the pendulum equation, $\ddot\theta+Y+\theta g/l=0$,  with an unspecified force $Y$, we find
\begin{equation}\dot S=(Ml^2/T)\dot\theta\, Y. \label{dot S}    \end{equation}
Since $Ml^2/T>0$, but not necessarily $\dot\theta$, the requirement  $\dot S>0$ confines the form for $Y$, of which the simplest is:  $Y=\alpha\dot\theta$,  $\alpha>0$. This force acts until  $\theta,\dot\theta=0$, implying  $E=\int T{\rm d}S$, or  $S=$ max.

Grains are not different in this respect. Because DEM possesses the proper dissipative forces that drive the system toward equilibrium, $S=$ max, it may treat grains as elementary. But if one needs to  {\it derive} 
dissipative terms setting up a continuum-mechanical theory, it is not clear how one can possibly avoid $S$. 

\subsection{Athermal Gas versus Thermal Solid}
Faced with the task to account for granular behavior, it may seem natural to always model the medium as a gas of  elementary grains. Yet, within the entropic context, the analogy between flying grains and gaseous atoms is not close: Energy  conservation holds only for atoms, not for elementary grains. Accordingly, thermodynamics and statistical mechanics work only for the former. 

ASM tries to ameliorate this by replacing the energy with 
volume or stress, taking the latter to be conserved. Yet energy conservation, related to time translational symmetry,  is a fundamental property of matter; constancy of volume or stress are merely experimental prescriptions.  

There are more problems:
Failure of the temperatures to equilibrate, invalidity of the fluctuation-dissipation theorem, jammed grains lacking the possibility to explore the phase space, ... All these in addition to the basic problem, $S\gg S_g$.
Now, if the gas model poses such difficulties, the conjecture that a grain is further away from an atom then a block of rock that has been sliced and diced, naturally arises.

Generally speaking, of the three phases of matter, the gaseous one is the simplest to describe. The solid phase became more easily accountable only after it was realized that, at low enough temperatures,  it may be modeled as a gas of free quasi-particles: mainly phonons, and in conductors, also free electrons. 
Considering a block of solid including them, the dissipated energy is not lost. 
With the total energy conserved, classical thermodynamics holds, and statistical mechanics (of phonons and electrons) is valid. Crucially, same holds for a stack of two blocks -- or a pile of grains, as long as they are macroscopic.

The notion that grains at rest are jammed, in need of shaking for phase space exploration, is now inappropriate: Phonons and electrons roam nearly as freely in a pile of grains as in a block of rock. The fluctuation-dissipation theorem, dealing with thermal fluctuations of phonons and electrons~\cite{kubo}, holds also in granular media. 

With granular microscopics reassuringly  healthy, one confidently proceeds to consider the 
macroscopic description of granular dynamics. A block of solid is typically an elastic medium. Cutting the block 
in half, with one part on top of the other, we expect them to again be elastic under shear if they do not slip. If they do,  we subtract the slipping portion from the total displacement to obtain the deforming one.  Further dicing the block to eventually arrive at many (macroscopic) pieces, the system is still elastic -- though we need to keep track of the deforming displacement. We call the associated strain field  elastic, denote it as $u_{ij}$, and use it to account for the coarse-grained elastic deformation of the system. 

Since the elastic stress $\pi_{ij}$ stems from the deformation of the grains, we have
$\pi_{ij}=\pi_{ij}(u_{kl})$. Or more completely,  $\pi_{ij}=\pi_{ij}(u_{kl},\rho)$.

Force equilibrium among grains is the condition for maximal entropy $S$ with respect to variation of  $u_{ij}$, see Sec.\ref{2-A}.  Therefore, any stable configuration of grains at rest is in equilibrium. Including phonons and electrons that  explore the phase space, {grains at rest are indeed in the most conventional of equilibria.}  

Off-equilibrium, granular dynamics is operative. To set up a continuum-mechanical theory, and derive its many dissipative terms, an entropic consideration that includes both $S$ and $S_g$ is necessary -- as has been done to derive GSH. 
Equilibrium is characterized by $S_{tot}=S+S_g$ being maximal, or approximately $S=$ max.  
Taking $S_g=$ max is correct only if $S=$ const. Yet since dissipation is ubiquitous among grains, and any dissipation heats up the grains, this is not at all a likely scenario.

\subsection{Thermal and Athermal Explanation of  Tapping} 

A heap of grains has many stable configurations, many local maxima of $S$. The logarithm of this number,  typically an increasing function of the density, is the Edward entropy, $S_{Ed}(\rho)$. Assuming it is $S_{Ed}(\rho)$ that is maximal in equilibrium implies that the density increases under tapping, simply because this  increases   $S_{Ed}$. 

Since  $S_g$ counts all granular states, in and off equilibrium, including the much more numerous ones with grains flying and jiggling, $S_{Ed}$ contains only a very small subclass of the states in $S_g$, or
\begin{equation}\label{ath3}
S_{Ed}\ll S_g\ll S.
\end{equation}
Drawing any conclusions from $S_{Ed}$ is justified only if $S$ and $S_g$ do not depend on the density. Yet we know they do.

The alternative explanation emplyong GSH involves $S$, rather than $S_{Ed}$. 
In a free column of grains, every layer (carrying the same load from above) has a given pressure. In GSH, as mentioned, the pressure is a function of the density and granular deformation, $P(\rho, u_{ij})\equiv\frac13\pi_{\ell\ell}$.
Under tapping, because grains periodically lose or loosen contact with one another, their deformation is slowly lost, $u_{ij}\to0$. As $u_{ij}$ relaxes, the density $\rho$ increases to maintain $P$. 
Underlying this train of arguments is an increase of $S$:  As $u_{ij}$ relaxes, the associated elastic energy dissipates into heat. This increase is incomparably larger than any change in $S_{Ed}$.  

Looking for an experiment to discriminate between both explanations, we note that the first  is independent of the pressure, while compaction results from constant pressure  in GSH. 
If grains are submerged in a liquid of the same density, gravitation does not produce any pressure. Starting from an initial stress, tapping will diminish $u_{ij}$, and with it also the stress, but will not increase the density.  

\begin{figure}[h]
\begin{center}\vspace{-1cm}
\includegraphics[scale=.35]{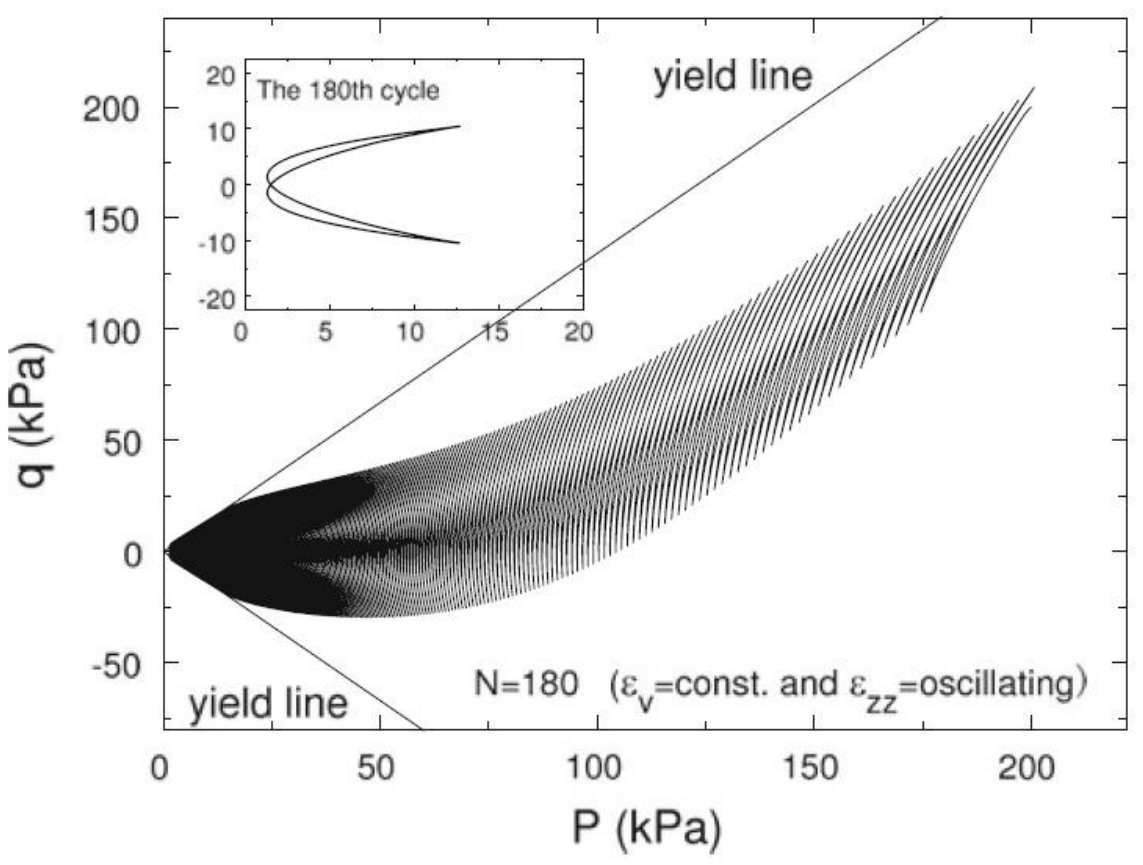}
\caption{GSH calculation depicting the relaxation
of the shear stress $q\equiv\sigma_{zz}-\sigma_{xx}$ and the pressure $P$,
in a triaxial geometry, for given 
shear oscillation, $\varepsilon_{zz}$, but no compression, $\varepsilon_{v}=$ const. 
(Inset amplifies the last calculated cycle.) }
\label{Fig1}
\includegraphics[scale=.35]{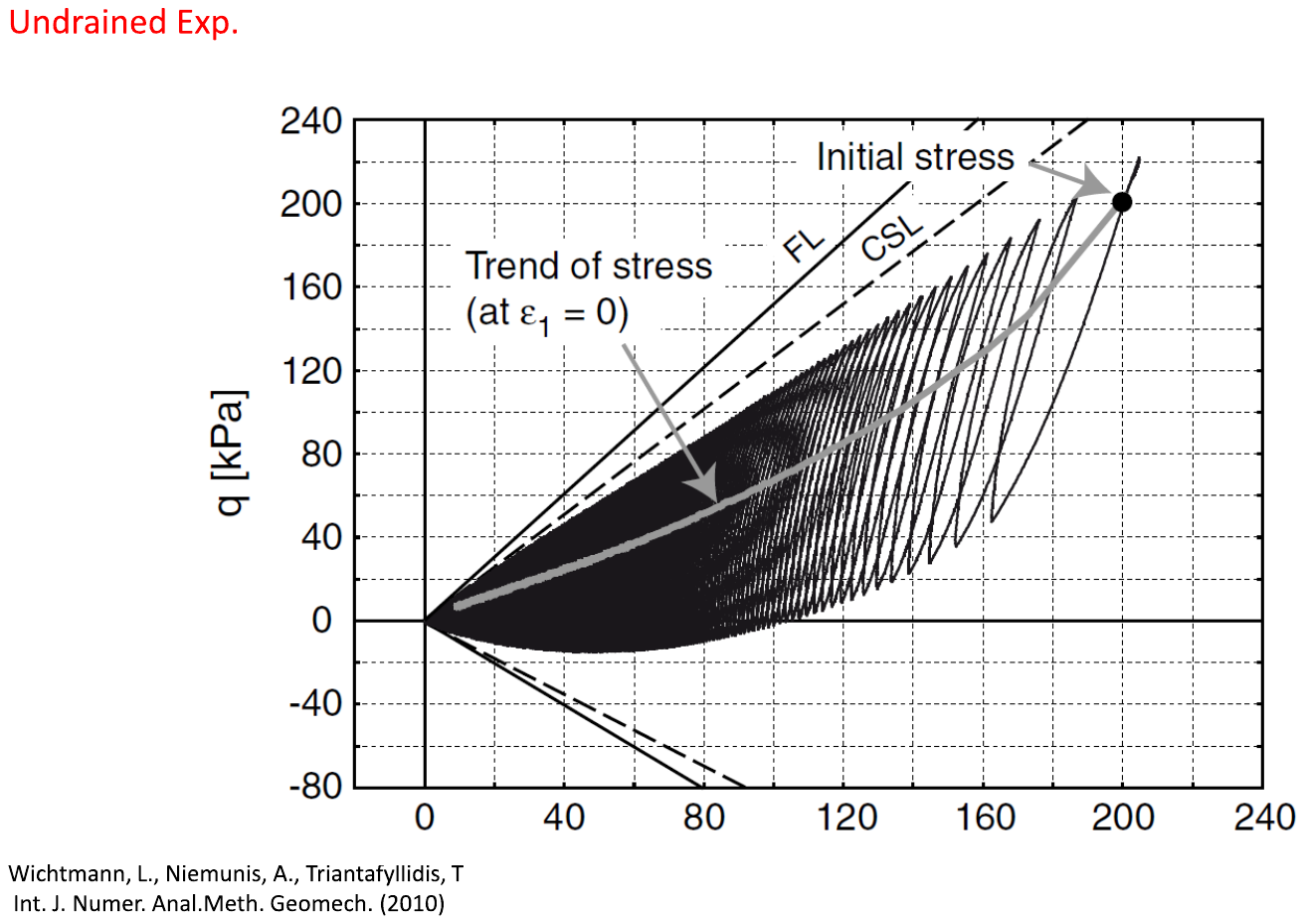}
\caption{Imposing isochoric deformation cycles to saturated sand in
undrained triaxial tests, Wichtmann et al.~\cite{Wichtmann09} obtained results qualitatively similar to that of Fig.\ref{Fig1}.}
\label{Fig2}
\end{center}
\end{figure}

According to GSH, small-amplitude cyclic shear jiggles grains and has essentially the same effect as tapping. 
Doing this at constant pressure, compaction takes place. But at constant volume, we only achieve $u_{ij}\to0$, causing  the pressure and shear stress to also vanish.  A GSH-calculation of the second case is rendered in Fig.\ref{Fig1}, the associated experiment is rendered in Fig.\ref{Fig2}. Details of the calculation may be found in~\cite{granRgudehus}, though its relevance to tapping was not realized then. 
See also appendix B for more results on compaction, both the reversible and irreversible branch, and the explanation of the memory effect.

\subsection{The Concerns of the Athermal Community} A large fraction of the granular physics community believes granular media cannot be treated by conventional tools of theoretical physics, because general principles,  including energy conservation, thermodynamics and the concept of equilibrium, are invalid. As should be obvious by now,  these are consequences of the athermal model, not abnormalities of granular media as such. 

Then there are those who do accept that equilibrium is given by $S=$ max, but  take this as a result of equilibrium statistical mechanics alone, lacking any relevance off-equilibria. This is a partial  and erroneous view. First of all, the entropy is also defined in local equilibrium and in generalized equilibrium, in which a few slowly relaxing variables are off their equilibrium values.  Starting off-equilibrium, the entropy grows continuously, by changing the value of slow variables, and by distributing conserved quantities, until it is maximal, and the system in equilibrium. For instance, a pendulum comes to a standstill hanging down, and the energy redistributes to achieve a uniform temperature. This evolution is accounted for by dissipative terms, which are derived by requiring that $S$ always increases -- as we did for the pendulum around Eq.(\ref{dot S}), and as was done setting up GSH.





\section{Granular Thermodynamics\label{sec2}}
\subsection{The Granular Temperature ${T_{g}}$\label{SecTg}}
In any uniform medium such as water or air, there are two length scales, macro- and microscopic. All degrees of freedom may be divided into either of these two groups. A hydrodynamic theory  takes the degrees of freedom from the first as explicit variables, each with an equation of motion. These including mass, momentum and energy density. 
Those from the second group are taken summarily, with their contribution to the energy lumped together as heat, and characterized by the temperature $T$. Irreversibility is caused by the macroscopic energy decaying into heat. 

In granular media, there is an intermediate,  mesoscopic group of degrees -- momentum and deformation of  individual grains. In DEM, these are explicit  variables. But for a hydrodynamic theory,  a summary inclusion again suffices, with their energy lumped into granular heat, quantified by $T_g$. Irreversibility is now caused by the macroscopic energy decaying into granular heat, and then on to true heat. This is what we term {\it two-stage irreversibilty}. 

We separate mesoscopic from microscopic degrees, with $T$ and $T_g$, instead of lumping them into one group and one temperature, not only because of the different length scales.  Equally important is the fact that $T_g$ is an independent state variable, on which the dynamics critically depends: The elastic stress relaxes when the grains jiggle, when $T_g\not=0$, while it is (within limits) independent of $T$. Lumping both into one temperature obscures this difference. 

Extending this argument, we see that a third temperature is superfluous. For instance, it is not useful to introduce a temperature for those degrees characterizing the surface roughness of grains. Though the length scale is distinctly smaller, no aspect of macroscopic granular dynamics depends critically on an associated temperature. These degrees are simply part of $T_g$. Similarly, it is not useful to introduce a configuration temperature $T_{Ed}$ associated with the Edward entropy, as it does not possess any significance independent from $T_g$, see also the discussion in Sec.\ref{athSED}. 

Both $T_g$ and $T$ are genuine temperatures, as each characterizes the energy of a group of degrees of freedom. Same holds for the granular and true entropy, $S_g$ and $S$: Each is the logarithm of the number of states in the associated group.  Any difficulties treating $T_g$ as a temperature arises only because $T$ is ignored. 
For instance, the granular temperatures $T_g$ of two systems in contact
are typically different --  seemingly a failure of  $T_g$ to equilibrate.
Given two granular systems, 1 and 2, with only 1 being excited, there are, in the steady state, four generally unequal
temperatures: $T^1,T_g^1,T^2,T_g^2$. And there are three ongoing energy fluxes: $(T^1_g\to T^1)$, $(T^1_g\to T_g^2)$, $(T_g^2\to T^2)$. 
This is in complete analogy to four conventional thermal systems, (1, 1a, 2, 2a), with only (1a) being heated by an external energy flux, and (1a,1), (1a, 2a), (2a, 2) in pairwise thermal contact.  We expect all four temperatures: $T_1,T_{1a},T_2,T_{2a}$ to be different.

Having stated this, we note a practical difference.
Taking the energy density as a function of the two entropy densities, $w=w(s, s_g)$, the conjugate variables are: 
\begin{equation}
T\equiv{\partial w}/{\partial s}, \qquad T_g\equiv{\partial w}/{\partial s_g}.
\end{equation}
Denoting $s_{tot}=s+s_g$, we may write
\begin{equation}\label{2-0}
{\rm d}w=T{\rm d}s+T_g{\rm d}s_g=T{\rm d}s_{tot}+(T_g-T){\rm d}s_g,
\end{equation} 
and identify $T{\rm d}s_{tot}$ as the equilibrium energy change for changes of the total entropy, and $(T_g-T){\rm d}s_g$ as the extra energy contribution for $T_g\not=T$. 
With $(T_g-T)$ characterizing the non-optimal energy distribution between the granular and microscopic degrees of freedom, the energy $w$ has a minimum at $T_g=T$, and is, expanded, given as: $w\propto(T_g-T)^2$. This also implies $(T_g-T)$ relaxes until it vanishes. 

Now, since $s\gg s_g$, and any granular motion at all occurs at  $T_g\gg T$, we have $T_g-T\approx T_g$,  $s_{tot}\approx s$,  and the rewriting of Eq.(\ref{2-0}) did not change anything. Therefore, we may simply take  $w\propto T_g^2$, with $T_g$ relaxing until equilibrium, $T_g=0$. 
And since $T$ does not  relax, we have $T=$ uniform in equilibrium, see more in Sec.\ref{2-A}.

A division into three scales works when they are well separated -- though this  is a problem of accuracy, not
viability. Scale separation is well satisfied in large-scaled, engineering-type experiments, less so in small-scaled
ones. Using glass or steel beads (typically larger) aggravates the problem. 
Nevertheless, when the system is too small for spatial averaging, one may still average over time and
runs to get rid of the fluctuations not contained in a hydrodynamic theory.

Finally, it is useful to realize that since granular heat $\int T_g{\rm d}s_g\propto T_g^2$ denotes the elastic and kinetic energy of the grains, it reduces,  in the limit of vanishing $\rho$,  in which only the kinetic part remains, to the energy or temperature of the kinetic theory $T_k$, implying  $T_k\propto T_g^2$. For the same reason, Eq.(\ref{Tg1}) for $T_g$ is, in  this limit,  the same as the energy balance of the kinetic theory.

\subsection{More on the Relation between $S_g$ and $S_{Ed}$\label{athSED}}
The Edwards entropy $S_{Ed}$, a function of the volume $V$, is employed with
\begin{equation}
{\rm d}S_{Ed}={\rm d}V/X
\end{equation}
as the basic thermodynamic relation for a {``\it mechanically stable agglomerate of infinitely rigid grains at rest"}~\cite{Edw}.
This {\em ansatz} is better appreciated by taking the granular entropy as $S_{g}(E,V)$ (while still neglecting phonons and electrons), writing
\begin{equation}\label {edwards entropy revisited}
{\rm d}S_{g}=(1/T_{g}){\rm d}E+(P/T_{g}){\rm d}V.
\end{equation}
For infinitely rigid grains at rest, as there is no kinetic or deformation energy, we
have $E\equiv0$, which remains zero however these grains are arranged,  ${\rm d}E\equiv0$, hence 
\begin{equation}
{\rm d}S_{g}=(P/T_{g}){\rm d}V\equiv{\rm d}V/X,
\end{equation}
reducing $S_g$ to $S_{Ed}$.
More generally, because grains are elastic and frequently in motion,  ${\rm d}E\not=0$, and Eq.(\ref{edwards entropy revisited}) holds. 

There is no independent Edward temperature $T_{Ed}\sim X$. It is simply what $T_g$ reduces to if one chooses to take grains as infinitely rigid and perennially at rest.

We also note that infinite rigidity is not a realistic limit in sand:
Because of the Hertz-like contact between
grains, very little material is deformed at
first contact, and the compressibility diverges at
the jamming point, for $u_{ij}\to0$, see Eq.(\ref{2b-2b}). This is a geometric fact
independent of the material's rigidity.

\subsection{The Elastic Strain Field ${u_{ij}}$ \label{Uij}}
A block of rock, if elastic,  is well accounted for by the elastic energy,  the elastic stress, and  the equation of motion of the strain $\varepsilon_{ij}$, 
\begin{align}\label{PartialEpsilon}
w=w(\varepsilon_{ij}),\quad \pi_{ij}=-\partial 
w/\partial\varepsilon_{ij},
\\\nonumber
\partial_t\varepsilon_{ij}=v_{ij}\equiv(\nabla_iv_j+\nabla_jv_i)/2.
\end{align}
Slicing the block in half, with one part on top of the other, we expect them to again be elastic under shear if they do not slip. If they do,  we 
subtract the slipping portion from the total displacement to obtain the deforming one. If one uses it to calculate the strain field, the first two above relations remain valid -- as only the deforming field leads to an elastic 
energy, which in turn leads to a restoring force that is the elastic stress. We call the new strain field elastic, denote it as $u_{ij}$,  and write
\begin{equation}\label{elaStress}
w=w(u_{ij}),\quad \pi_{ij}(u_{ij})=-\partial w/\partial u_{ij}.
\end{equation}
Further cutting the block to eventually arrive at many macroscopic pieces, this consideration still holds. We again take the portion of the strain that deforms the grains 
as $u_{ij}$, and again, only this field senses the restoring stress. 

An analogy should make the last half sentence clearer. The wheels of a car driving up a slippery slope have a gripping portion  $\theta_{g}$, and a slipping portion $\theta_{s}$. The force on the car is $\partial w_g/\partial\ell$, with $w_g$ the gravitational energy and $\ell$ the distance traversed, implying that the torque on the wheels is  $\partial w_g/\partial\theta_{g}$.

Eqs.(\ref{elaStress}) are useful in two aspects.  First, given an evolution equation for $u_{ij}$ (that is yet to be found, see appendix A.1), we have a granular theory as mathematically complete as ideal elasticity. And there is no need to consider, in addition, the plastic strain rate $\partial_t\varepsilon_{ij}^{p}\equiv\partial_t\varepsilon_{ij}
-\partial_tu_{ij}$. 
Second, given the elastic stress-strain relation $\pi_{ij}=\pi_{ij}(u_{kl})$,  any information on $\pi_{ij}$ is equivalent to that on $u_{ij}$. Only the first is  directly measurable. Note for given $\rho$,  the stress $\pi_{ij}\equiv-\partial w/\partial u_{ij}$ is a monotonic function of $u_{ij}$, as any stable energy is convex, $\partial^2 w/\partial u_{ij}\partial u_{k\ell}>0$. 

In soil mechanics, the plastic strain rate is related to flow rules and deemed important. Moreover, the energy is often taken as a function of both $u_{ij}$ and $\varepsilon_{ij}^{p}$~\cite{Houlsby2}. Yet since granular energy is either  elastic or kinetic, stored either in the deformation or motion of the grains, it is not clear what the energy contribution of  $\varepsilon_{ij}^{p}$ is.

Finally, we note that the elastic strain is not obtained by coarse-graining its mesoscopic counterpart, $u_{ij}\not=\langle u_{ij}^{mes}\rangle$. The reason is, both the energy and stress are given by averaging:  $w=\langle w^{mes}\rangle$,  $\pi_{ij}=\langle\pi_{ij}^{mes}\rangle$. Since  
${\rm d}w=\langle{\rm d}w^{mes}\rangle=-\langle \pi_{ij}^{mes}{\rm d}u_{ij}^{mes}\rangle=-\langle \pi_{ij}^{mes}\rangle{\rm d}u_{ij}$ and $\langle \pi_{ij}^{mes}{\rm d}u_{ij}^{mes}\rangle\not=\langle \pi_{ij}^{mes}\rangle{\rm d}\langle u_{ij}^{mes}\rangle$, we have ${\rm d}u_{ij}\not={\rm d}\langle u_{ij}^{mes}\rangle$.

\subsection{ Granular Equilibrium Conditions\label{2-A}} 

The state variables of any granular media are the density $\rho$, the
momentum density $\rho v_i$, the two entropy densities $s,s_g$,  and the elastic strain $u_{ij}$. Denoting the energy density in the rest frame ($v_i=0$) as $w=w(s, s_g, \rho, u_{ij})$, with $T\equiv{\partial w}/{\partial s}$, $T_g\equiv{\partial w}/{\partial s_g}$, $\mu\equiv{\partial w}/{\partial\rho}$ and $\pi_{ij}\equiv-{\partial w}/{\partial u_{ij}}$,  we have
\begin{equation}\label{2-1}
{\rm d}w=T{\rm d}s+T_g{\rm d}s_g+\mu{\rm d}\rho-\pi_{ij}{\rm d}u_{ij}.
\end{equation} 
Another useful conjugate variable, the fluid pressure $P_T$, belongs to  the energy density per unit  mass, $w/\rho$, and is given (with $V$ denoting the volume) as
\begin{equation}\label{pressure}
P_T\equiv-w+sT+s_gT_g+\mu\rho=-\frac{\partial(wV)}{\partial V}=-\frac{\partial(w/\rho)}{\partial(1/\rho)}.
\end{equation}
In dry granular media, $P_T$ is the pressure exerted by jiggling grains, see Eq.(\ref{2b-5-2}).

Equilibrium conditions,  valid irrespective of the energy expression  $w(s, s_g, \rho, u_{ij})$, are
obtained by formerly requiring $S=\int s\,{\rm d}^3r=$ max, for given energy $\int w\,{\rm d}^3r$ and mass $\int \rho\, {\rm d}^3r$, with $T_g$ allowed to relax.  Employing Eq.(\ref{2-1}), we first obtain 
\begin{equation}
\nabla_iT=0,\quad T_g=0.
\end{equation}
%
Usually, $T_g$ vanishes quickly. After this has happened,  $d\rho$ and $du_{\ell\ell}=-d\rho/\rho$ no longer vary independently. They therefore share a  solid-like equilibrium condition,  
\begin{equation}\label{2a-1}
\nabla_i(\pi_{ij}+P_T\delta_{ij})=\rho\, {g}_i.
\end{equation}
This is the expression of force equilibrium for the jammed state. 

If $T_g$ is kept finite, say by tapping, the system may further increase its entropy by independently varying $\rho$ and $u_{ij}$. It then arrives at two fluid-like equilibrium conditions, the first of which  requires the shear stress to vanish, and a free surface to be horizontal,
\begin{equation}\label{2a-2} \pi_{ij}=0, \quad
\nabla_iP_T=\rho\, {g}_i. \end{equation}

\section{Conclusions}
As grains are too large to display thermal fluctuations,  they are widely taken as particles without any internal structure, and  considered ``{\it athermal.}'' Though an excellent approximation for DEM and the kinetic theory, it fails in any entropic considerations, because the inner-granular,  microscopic degrees of freedom are then the dominating ones. We have clarified the reasons why and when an athermal approach works, and when it fails. 

Embracing the notion that the macroscopic theory for granular media is that of a thermal solid that has been sliced and diced, and introducing two temperatures,  true and granular, one arrives at GSH, a comprehensive yet simple hydrodynamic theory capable of accounting for a wide range of granular phenomena. 


As this paper serves a dual purpose, to deflect concerns people accustomed to the athermal model have, and to ease their introduction to GSH, we present a minimalist version of GSH in appendix A, complete with a number of analytic solutions. (The original version, more complex and realistic, was derived a decade ago~\cite{granR1}, and has since been employed to account for many experiments~\cite{granR4,granRexp}.) 

A summary of this version of GSH is given here, to display its mathematical structure. Explanations are found in appendix A.
The state variables of any granular system are the density 
$\rho$, the momentum density $\rho v_i$, the granular entropy density $s_g$,  and the elastic strain $u_{ij}$, see Sec.\ref{2-A}. (The entropy $s$ is excluded here, as we do not aim to consider effects such as temperature diffusion or thermal expansion  -- though $s$ was of course included in deriving GSH.) 
Their evolution equations are: the continuity equation, momentum balance, balance of the granular entropy, and the evolution equations for the elastic strain. Denoting $\Delta\equiv-u_{kk}$, $u^*_{ij}\equiv u_{ij}+\frac13\Delta$, $v_{ij}\equiv\frac12(\nabla_iv_j+\nabla_jv_i)$ as the strain rate, and $v^*_{ij}$ again traceless, we have 
\begin{align}\label{GSH16}
\partial_t&\rho+\nabla_i(\rho v_i)=0,
\\
\partial_t&(\rho v_i)+\nabla_i(\sigma_{ij}+\rho v_iv_j)=\rho g_i,
\\
&\,\sigma_{ij}=\pi_{ij}+P_T\delta_{ij}-\eta_1T_gv^*_{ij},
\\
&\,\pi_{ij}(u_{ij})\equiv-{\partial w\,\,}/{\partial u_{ij}},\,\,
P_T\equiv-{\partial(w/\rho)}/{\partial(1/\rho)},
\\
\partial_t&T_g=-R_T[T_g(1-\xi_T^2\nabla^2_i)T_g-f^2v^*_{ij}v^*_{ij}],
\\
\partial_t&u^*_{ij}=v^*_{ij}-\lambda T_gu^*_{ij},
\\\label{GSH22}
\partial_t&\Delta+v_{\ell\ell}=\alpha_1u^*_{ij}v^*_{ij}-\lambda_1 T_g\Delta.
\end{align}
(Note that $\sigma_{ij}$  is fixed if $w$ is given. In constitutive models, it is typically taken as an expression with little constraints, to be extracted from data alone.)
This set of partial differential equations is closed if the energy density $w$ and the seven transport coefficients: 
$\lambda,\lambda_1,\alpha_1,R_T,\xi_T,f,\eta_1$ are known [$f$ is not independent, cf~Eq.(\ref{ath-f})]. These scalars provide the leeway GSH has for fitting experiments, from elasto-plastic motion to fast dense flow. 
A simplified expression for $w$ is given in Eqs.(\ref{ath26},\ref{2b-2},\ref{eq1}). The transport coefficients, all functions of the density,  are taken as constant here.

\newpage
\appendix
\section{The Expressions of GSH}
We provide a brief derivation of  Eqs.(\ref{GSH16}-\ref{GSH22}) in this section.
\subsection{Relaxation equation for ${u_{ij}}$\label{EqMoU}}
First, we consider what $\partial_tu_{ij}$ is: The free surface of a granular system at rest is frequently tilted. When perturbed, when the grains jiggle and $T_g\not=0$, the tilted surface will
decay and become horizontal. The stronger the grains jiggle, the faster the decay is. We take this as indicative of a system
that is {\it elastic for $T_g=0$, transiently elastic for $T_g\not=0$,
with a stress relaxation rate that increases with $T_g$}. 
We take the rate $\propto T_g$, as this yields rate-independence for the elasto-plastic regime. 

A relaxing stress is typical of any visco-elastic system such as polymers~\cite{polymer-1,polymer-2,polymer-3,polymer-4}. 
In granular media, the relaxation rate is not a material constant, but a function of the state variable $T_g$ -- a behavior that we call {\em variable transient elasticity}. 
Remarkably, by taking the evolution equation for $u_{ij}$ to reflect {\it variable transient elasticity}, one finds that the result suffices to capture elasto-plasticity. 

For the traceless part $u^*_{ij}$, we start with the equation of elasticity, then add a relaxation term $\propto T_g$. The next line is a rewrite that clarifies the stationary state,
\begin{align}\label{eqU2} 
\partial_tu^*_{ij}&=v^*_{ij}-\lambda T_gu^*_{ij}\qquad (\text{with}\,\,u_c\equiv1/\lambda f)
\\
&=-\lambda T_g[u_{ij}^*-u_c\frac{v_{ij}^*}{v_s}\frac{fv_s}{T_g}], 
\nonumber
\end{align} 
where $v_s\equiv\sqrt{v^*_{ij}v^*_{ij}}>0$. 
The negative trace of the elastic strain, $\Delta\equiv-u_{kk}$, obeys a slightly more complicated  equation, with a term preceded by $\alpha_1$ accounting for dilatancy. With $u_s\equiv\sqrt{u^*_{ij}u^*_{ij}}>0$, it is 
\begin{align}\label{eqD2}
\partial_t\Delta+v_{\ell\ell}&=\alpha_1u^*_{ij}v^*_{ij}-\lambda_1 T_g\Delta\qquad (\text{with}\,\,\Delta_c/{u_c}\equiv{\alpha_1}/{\lambda_1f})
\\\nonumber
&=-\lambda_1T_g\left[\Delta-\Delta_c\frac{u_s}{u_c}\frac{fv_s}{T_g}\frac{v^*_{ij}}{v_s}\frac{u^*_{ij}}{u_s}\right].
\end{align}
As we shall see, in the rate-driven stationary state, we have $fv_s=T_g$, $u_s=u_c$, and  ${v^*_{ij}}{u^*_{ij}}=v_su_s$, reducing the right hand side to $-\lambda_1T_g\left[\Delta-\Delta_c\right]$.

If the anisotropy of the stress changes appreciably, convective terms (or the objective time derivative) become important, then one needs the substitution~\cite{polymer-1}
\begin{equation}
\partial_t u_{ij}\to(\partial_t+v_k\nabla_k)u_{ij}
+\Omega_{ik}u_{kj}-u_{ik}\Omega_{kj},
\end{equation}
where $\Omega_{ik}\equiv\textstyle\frac12(\nabla_iv_k-\nabla_kv_i)$.

\subsection{The Elastic Energy ${w_\Delta(u_{ij},\rho)}$ \label{Wela}}
As a first step to specify   $w=w(s, s_g, \rho, u_{ij})$, we note that due to the lack of interaction among 
grains, $w$ vanishes when the grains are neither deformed nor jiggling. Taking
\begin{equation}\label{ath26}
w(\rho,s_g,u_{ij})=w_\Delta(\rho,u_{ij})+w_T(\rho,s_g),
\end{equation}
we therefore require $w_T\to0$ for $s_g\to0$, and  $w_\Delta\to0$ for $u_{ij}\to0$. 
The dependence on the true entropy density $s$ is not specified, because we are not 
at present dealing with any temperature related phenomena. But this is easily ameliorated if needed.

The elastic energy $w_\Delta$ needs to be chosen such that it yields the right static stress distribution and elastic wave propagation.  
Equally important,  it needs to account for the fact that elastic solutions are untenable if (1)~the density is too small and the grains loose contact with one another, or if (2)~the shear stress is too large, say when a slope is too steep, and the grains start slipping. Because both instabilities may happen in equilibrium, when grains are at rest,  they have to be encoded in the energy, not in evolution equations or transport coefficients. This is done most easily by a transition of the energy from convex to concave at the instabilities, because elastic solutions are stable only if the energy is convex. 
Defining the random loose packing density $\rho_{lp}$ as the minimal one sustaining an elastic solution, we require the energy to possess a convexity transition at  $\rho_{lp}$, such that no elastic solution is stable for $\rho<\rho_{lp}$. 
Similarly, we define a yield stress $\pi_{yie}$ such that the energy is, at given pressure $P$, convex for ${\pi_s}\equiv\sqrt{\pi^*_{ij}\pi^*_{ij}}<\pi_{yie}$ and concave for ${\pi_s}>\pi_{yie}$. 

This implies that the angle of stability $\varphi_{st}$ is -- for an idealized, uniform and infinite slope -- given by  $\tan\varphi_{st}\equiv{\pi_{yie}}/\sqrt2P$: 
On an infinite plane inclined by the angle $\varphi$,  with $y$ the depth of the granular layer on the plane, and $x$ along the slope, we take the stress to be $\pi_{xx},\pi_{yy},\pi_{zz}=P_\Delta$, $\pi_{xy}=\pi_s/\sqrt2$, $\pi_{yz},\pi_{xz}=0$. Integrating $\nabla_j\pi_{ij}=g_i\rho$ assuming a variation only along $y$, we find $\pi_{xy}=g\sin\varphi\int\rho(y)dy$ and $\pi_{yy}= \pi_{xy}/\tan\varphi$. The angle of stability $\varphi_{st}$ is reached when the energetic instability of Eqs.(\ref{2b-3}) is breached. 
(Effects from proximity to the wall or floor, such as clogging, or the fact that the angle increases when the layer is very thin,  are not considered here.) The smaller angle of repose $\varphi_{re}$ -- at which any granular flow first comes to an halt -- is a property of granular dynamics, and as discussed in appendix D.2, given by $\tan\varphi_{re}={\pi_c}/\sqrt2P$, with $\pi_c$ the critical shear stress and $\pi_c/P$ the critical friction.

A simple expression satisfying all the above points is, with  ${\cal A,B}>0$,
\begin{align}
\label{2b-2}  
w_\Delta=\sqrt{\Delta }&[2 {\mathcal B}(\rho) \Delta^2/5+ {\mathcal A}(\rho)u_s^2],
\\\label{2b-2b} P_\Delta=\sqrt\Delta&({\cal B}\Delta+{\cal A}{u_s^2}/{2\Delta}),\,\, \pi_{ij}^*=-2{\cal A}\sqrt\Delta\, u_{ij}^*,
\\\label{2b-1}
{4P_\Delta}/{\pi_s}&=2({\cal B}/{\cal
A})(\Delta/u_s)+{u_s}/{\Delta},\qquad\qquad     
\end{align} 
where $P_\Delta\equiv\pi_{\ell\ell}/3$. 
These expressions have been validated for: 
(1)~static stress distributions in silo, sand pile,
point load on a granular sheet, calculated employing $\nabla_i\pi_{ij}=\rho g_i$, see~\cite{granR1,exp1.,exp2.};
(2)~incremental stress-strain relation from varying static stresses~\cite{exp3.};
(3)~propagation of elastic waves at varying static stresses~\cite{exp5.}.

Moreover, the energy is convex only for
\begin{equation}\label{2b-3}
u_s/\Delta\le\sqrt{2{\cal B}/{\cal A}},\quad\text{or}\,\,\, \pi_s\le\pi_{yie}\equiv P_\Delta\sqrt{2{\cal A}/{\cal B}},
\end{equation} 
turning concave if this condition is violated. Because $\pi_{yie}/P_\Delta$ is observed to be not or only weakly density dependent, 
we take ${\cal B}/{\cal A}=$ const, 
and  assume   
\begin{equation}\label{2b-4} {\cal B}={\cal B}_0
\left[\frac{\rho-\bar\rho}{\rho_{cp}-\rho}\right]^{0.15},
\end{equation}
where ${\cal B}_0>0$ is a constant,  
and $\bar\rho\equiv\frac19(20\rho_{\ell p}-11\rho_{cp})$. 
($\rho_{cp}$ is the {\it random-close packing density}, the highest one at which grains may remain uncompressed. For lack of space, grains cannot rearrange at $\rho_{cp}$, diminishing  plasticity. Note $\bar\rho<\rho_{\ell p}<\rho_{cp}$, with  $\rho_{cp}-\rho_{\ell p}\approx\rho_{\ell p} -\bar\rho$.)
This expression is constructed to account for four granular characteristics:
(1)~It is concave for any density smaller than $\rho_{\ell p}$, such that no elastic state is stable. 
(2)~It is convex between $\rho_{\ell p}$ and $\rho_{cp}$, ensuring the stability of elastic solutions in this region. (3)~The density dependence of sound velocities (as measured by Hardin and Richart~\cite{hardin}) is well rendered by  $\sqrt{{\cal B}/\rho}$. 
(4)~The slow divergence at
$\rho_{cp}$ approximates the fact that the system is much stiffer for  $\rho=\rho_{cp}$. 


The above expressions may be amended for more realism. First, including the third invariant, $u_3\equiv
\sqrt[3]{u^*_{ij}u^*_{jk}u^*_{ki}}$, see~\cite{exp7.}, $u_{ij}$ and $\pi_{ij}$ are no longer colinear, $u_{ij}^*/u_s\not=\pi_{ij}^*/\pi_s$, though they still share the same principal axes. As a result, Coulomb and Lade-Duncan yield laws may be accounted for -- in addition to the Drucker-Prager one.  
Second, including higher order strain terms allows one to account for compressional instabilities, the fact that a granular state can only sustain a maximal static pressure~\cite{granR2,granR4}. 

\subsection{Granular Heat ${w_{T}(s_g,\rho)}$\label{WT}}
In this section, we specify granular heat ${w_{T}(s_g,\rho)}$, and derive the fluid pressure $P_T$ from it.  
In the gaseous phase, grains have only kinetic energy. With $n=\rho/M$ the number of
grains per unit volume (and $M$ the average mass of the grain), the energy density is $w_T=\frac32T_kn$.  
($T_k$ is the granular temperature for granular gas.) 
At lower granular temperatures, when enduring contacts dominate, this formula is invalid. Going to the small-temperature limit, we expand $w_T$ in $s_g$, requiring it to be minimal for $T_g=0$, see the consideration in Sec.\ref{SecTg}, below Eq.(\ref{2-0}), 
\begin{equation}\label{eq1}
w_T=\frac{s_g^2}{2\rho b(\rho)}=\frac{\rho b}2T_g^2,\quad T_g\equiv\frac{\partial w_T}{\partial s_g}=\frac{s_g}{\rho b}.
\end{equation}
Assuming only analyticity of $w_T$ (ie. that it may be expanded in a power series),  this expression is force-independent and fairly general. 

Instead of looking for a formula interpolating between $T_g$ and $T_k$, we employ Eq.(\ref{eq1}) for all values of $T_g$. 
For large $T_g$, this means we identify $w=\frac12\rho b{T_g^2}$ with
$w=\frac32T_k\rho/M$, or $T_k=bMT_g^2/3$.  
Of course, only one temperature can be the actual temperature, which equilibrates with the true temperature $T$ of the grains. Strictly, it is  $T_k$ in the gaseous, and $T_g$ in the solid phase. Yet pragmatically,  equilibration occurs only in the solid phase, for vanishing $T_g$. 
In the gaseous phase, equilibration would falsely imply the existence of super hot grains of more than million degrees. 

Eq.(\ref{eq1}) works well: For instance, in the kinetic theory, the pressure is found $\propto T_k$, 
see~\cite{luding2009,Bocquet}, while it is $\propto T_g^2$ in GSH,
\begin{equation}\label{2b-5-2}
-P_T\equiv\left.\frac{\partial (w_T/\rho)}{\partial 1/\rho}\right|_{s_g}=\left.\frac{\partial [(w_T-T_gs_g)/\rho]}{\partial 1/\rho}\right|_{T_g}=\frac{T_g^2\rho^2}{2}\frac{\partial b}{\partial\rho}.
\end{equation} 
We choose  $b=b(\rho)$ such that it is $P_T\propto w_T$ for  $\rho\to0$, and $P_T\propto w_T/(\rho_{cp}-\rho)$ for $\rho\to\rho_{cp}$, see~\cite{luding2009,Bocquet},
\begin{equation}\label{2b-5}
b=\frac{b_1}\rho+b_0\left[1-\frac{\rho}{\rho_{cp}}\right]^a,\quad
P_T=\frac{T_g^2\rho^2}{2}\left[\frac{b_1}{\rho^2}+\frac{b_0}{(\rho_{cp}-\rho)^{1-a}}\right]
\equiv g_pT_g^2, \end{equation}
with $a\approx0.1$. [The quantity $g_p$ in Eqs.(22) of~\cite{granRexp} contains a misprint.]
Given $ {\cal B}(\rho)$ of Eq.(\ref{2b-4}), there is also a contribution $\propto\Delta^{2.5}$ to $P_T$ from  $w_\Delta$. It is neglected because it is much smaller than the elastic one, $P_\Delta\propto\Delta^{1.5}$ for  $\Delta$ small.

\subsection{Relaxation Equation for ${T_g}$ \label{RelaxTg}}
Starting from the balance equation for $s_g\propto T_g$, we can easily write down the equation for $T_g$, which is  increased by viscous heating $(fv_s)^2$ and decreased by relaxation $\propto T_g^2$,
\begin{equation}
\label{Tg1} 
\partial_tT_g/R_T=-T_g(1-\xi_T^2\nabla^2_i)T_g+(fv_s)^2+T_e^2.
\end{equation}
The relaxation rate $R_TT_g$ is of order $10^3$/s in dense media; the diffusion length $\xi_T$ is a few granular diameters. For $v_s=0$ and $T_g$ uniform, we have, in the steady state, $T_g=T_e$, where $T_e$ is the granular temperature produced by external perturbations, say a sound field or tapping. We take $T_e=0$ here, but shall return in appendix B to discuss its significance. 
For granular gas,  $T_g\propto\sqrt{T_k}$, Eq.(\ref{Tg1}) reduces to the energy balance of the kinetic theory. 
For steady and  uniform $v_s$, we quickly arrive at 
\begin{equation}\label{TgVs}
T_g= f v_s.
\end{equation}
With $\eta_1T_g$ the shear viscosity, see Eq.(\ref{cauchy stress}), we have (cf.~\cite{granR4})
\begin{equation}\label{ath-f}
f\equiv\sqrt{\eta_1/R_Tb\rho},
\end{equation}
which is therefore not independent. 

For fast dense flow, 
vigorous agitation of the grains reflects a large $T_g$. For slow, elasto-plastic motion, $T_g$ is too small to be directly observable. Aside from an occasional slip, grains  
essentially participate in the macroscopic motion of the given shear rate, with no perceptible deviations that would contribute to $T_g$. Nevertheless, such a slip leads 
to vibrations of the gains, implying a $T_g$ still many orders of magnitude larger than the true temperature $T$. As this changes the elastic property by enabling 
stress relaxation, $T_g$ remains relevant in the elasto-plastic regime.

\subsection{Mass and Momentum Conservation \label{sigma}}

Mass and momentum conservation, 
\begin{equation}\label{MassMomentum}
\partial_t\rho+\nabla_i(\rho v_i)=0,\qquad
\partial_t(\rho v_i)+\nabla_i(\sigma_{ij}+\rho v_iv_j)=\rho g_i,
\end{equation}
are part of GSH; they are to be solved in conjunction with  Eqs.(\ref{eqU2},\ref{eqD2},\ref{Tg1}), the evolution equations for  $\Delta, u^*_{ij}$and $T_g$. 
The  total (or Cauchy) stress is
\begin{equation}\label{cauchy stress}
\sigma_{ij}=\pi_{ij}+P_T\delta_{ij}-\eta_1T_gv^*_{ij},
\end{equation}
with the elastic stress $\pi_{ij}$ from Eq.(\ref{2b-2b}) and the fluid pressure $P_T\propto T_g^2$ from Eqs.(\ref{2b-5}). The third term is the viscous stress, with the shear viscosity $\eta_1T_g$. 

\subsection{Summary} 
The variables of GSH are: $u_{ij},T_g,\rho,v_i$, their evolution equations  form a closed set of nonlinear, partial differential equations, which need to be solved for a range of boundary conditions. Below, we shall be solving it in various simple limits.

The statics of GSH  is given by the elastic energy $w_\Delta$, with the two coefficients, $\cal A,B$, and the thermal energy $w_T$, with the coefficient $b$. Their density dependence is specified in Eqs(\ref{2b-4},\ref{2b-5}). The dynamics consists of the Cauchy stress, $\sigma_{ij}$ of Eq.(\ref{cauchy stress}), and the  relaxation equations for $T_g$, $u^*_{ij}$ and $\Delta$, which share six transport coefficients: 
\begin{equation}
\Delta_c,\, \lambda,\, \lambda_1,\, R_T,\, \xi_T,\, \eta_1,
\end{equation}
all functions of $\rho$, left unspecified here. (With $f\equiv\sqrt{\eta_1/R_Tb\rho}$ and $u_c\equiv1/\lambda f$, neither is independent.) 
In evaluating GSH, we take  $\rho=$ const, such that all coefficients also are. This renders a number of analytic solutions possible. Frequently, experiments are performed at constant pressure P, or a stress component. 
Then the density may change, and the coefficients with it. This algebraically more involved case usually need to be solved numerically, something we shall eschew in this paper.

The stationary solutions of $u^*_{ij}$ and $\Delta$, see Eqs.(\ref{eqU2},\ref{eqD2}), are respectively 
\begin{equation}
\frac{u_s}{u_c}=\frac{fv_s}{T_g},\quad
\frac{\Delta}{\Delta_c}=\frac{u_s}{u_c}\frac{fv_s}{T_g}.
\end{equation}
Employing in addition  $\frac{\pi_s}{\pi_c}=\frac{\sqrt\Delta\, u_s}{\sqrt{\Delta_c}\, u_c}$ from~Eq.(\ref{2b-2b}), we have 
\begin{equation}\label{allProp}
\frac{\pi_s}{\pi_c}=\frac\Delta{\Delta_c}=\frac{u_s^2}{u_c^2}=\frac{(fv_s)^2}{T_g^2},\quad
\frac{\pi_{ij}^*}{\pi_s}=\frac{u_{ij}^*}{u_s}=\frac{v_{ij}^*}{v_s}.
\end{equation}
In addition to $T_g= f v_s$, see  Eq.(\ref{TgVs}), we have $u_s=u_c$, $\Delta=\Delta_c$, and the critical stress, 
\begin{equation}\label{cs}
\pi_c=\pi_s(\Delta_c,u_c),\quad P_c=P_\Delta(\Delta_c,u_c).
\end{equation}
This is the state in which all three variables are stationary, with $\Delta_c,u_c,P_c,\pi_c$ rate-independent -- ie. possessing values that are independent of the shear rate $v_s$. 

Driving a granular system at a slow shear rate $v_s$, the system executes the rate-independent elasto-plastic motion, and the stress is given by the  first term of Eq.(\ref{cauchy stress}), $\sigma_{ij}=\pi_{ij}$. This is where the hypoplastic model holds. In steady state, we have the critical stress, $P=P_c$ and $\sigma_s=\pi_c$.
At higher rates, the next two terms can no longer be neglected, and  we enter the regime of fast dense flow and the $\mu(I)$-rheology. For steady flow, with  $P_T\propto T_g^2=(fv_s)^2$ and  $\eta_1T_gv^*_{ij}=\eta_1fv_sv^*_{ij}$, we have, 
\begin{equation}\label{27}
P=P_c+e_p(\rho)v_s^2,\quad \sigma_s=\pi_c+e_s(\rho)v_s^2.
\end{equation}
Being  quadratic, the corrections $\sim v_s^2$ come on slowly, leaving the rate-independent regime,  $\sigma_{ij}=\pi_{ij}$, to persist for many orders of magnitude in the rate.

In studying the relaxation dynamics, we may differentiate between two cases. First, driving a granular system at a given shear rate, with $T_g$ quickly settling into the stationary solution, $T_g=fv_s$, the two remaining relaxation Eqs.(\ref{eqU2},\ref{eqD2}) are
\begin{align}\label{sta1} 
\partial_tu^*_{ij}&=v^*_{ij}-\lambda fu^*_{ij}v_s,
\\\nonumber
\partial_t\Delta+v_{\ell\ell}&=\alpha_1u^*_{ij}v^*_{ij}-\lambda_1\Delta fv_s.
\end{align} 
Second, to study the dynamics at given stress, or equivalently, at given elastic strain, $\partial_tu_s,\partial_t\Delta=0$, we insert Eq.(\ref{allProp}) into  Eq.(\ref{Tg1}) to find
\begin{align}\label{sta2}
\partial_tT_g&/(R_TT_g)+\xi_T^2\nabla^2_iT_g=-T_g(1-\pi_s/\pi_c), \\ fv_s&=T_g\sqrt{\pi_s/\pi_c}.
\nonumber\end{align}
Both equations are very useful. 
The rate-independent Eqs.(\ref{sta1}) account for the approach to the critical state at given rate, and for load/unload. It possesses the same mathematical structure as the hypoplastic model~\cite{kolymbas1,kolymbas2}, and yields very comparable results for elasto-plastic motion.
Equations.(\ref{sta2}), on the other hand,  describe creep, angle of repose, shear band, and it reduces to Kamrin's nonlocal constitutive relation~\cite{kamrin,kamrin2} for stationary flows, $\partial_tT_g\propto\partial_tv_s=0$.   

\section{Tapping and Compaction\label{D}}
Numerous experiments employing varying external perturbations show a compaction of the packing of grains, see the review article~\cite{1nico}. 
This phenomenon is widely accounted for by  employing the Edwards 
entropy $S_{Ed}$, or some generalization of it. We have reasons to doubt its basic assumption: Considering $S_{Ed}$ implies that all other, much larger entropy contributions remain constant, see the discussion in the introduction, also at the end of Sec.\ref{SecTg}. 
Moreover, this understanding makes compaction a singular effect, in need of a special treatment not apparently useful for any other granular phenomena. 
In contrast, we believe that compaction is well-embedded into other granular phenomena, and closely related to an observation familiar to engineers --  {\it density  increase at given pressure under cyclic shear. }
The point is, shear rates produce $T_g$, and $\Delta$ relaxes if $T_g\not=0$, see~Eq.(\ref{eqD2}). Since the pressure $P_\Delta={\cal B}(\rho)\Delta^{1.5}$, see  Eq.(\ref{2b-2b}), is kept constant, the density $\rho$ increases to compensate. 
(We neglect the deviatory stress for simplification.)

The effect of compaction remains the same however $T_g$ is  generated, via a shear rate $fv_s$, or by an external perturbation $T_e$ (generated by sound field, periodic water injection, or tapping), because including the external perturbation $T_e$, the uniform, stationary solution of Eq.(\ref{Tg1})  is 
\begin{equation}\label{TE}
T_g^2={(f v_s)^2+T_e^2}.
\end{equation}

\subsection{Reversible and Irreversible Compaction}
We consider the total pressure $P=P_\Delta+P_T$, see Eqs.(\ref{2b-2b},\ref{2b-4},\ref{2b-5},\ref{cauchy stress}),  
\begin{equation}\label{gc5}
P_\Delta={\cal B}(\rho)\Delta^{1.5},\quad P_T= g_p(\rho)T_g^2,
\end{equation}
with $\cal B$ and $g_p$ monotonically increasing functions of $\rho$. 
For $T_g$ small, the seismic pressure $P_T$ may be neglected, and $\rho$ increases as $\Delta$ relaxes,  because $P=P_\Delta=$ const. This portion of the compaction is irreversible. Most soil-mechanical experiments are in this limit. 

For  larger $T_g$, the seismic pressure $P_T$ cannot be neglected. So $\Delta$ relaxes with the density increasing irreversibly and $P_\Delta+P_T=$ const. Since $g_p$ is a more sensitive function of $\rho$ than  $\cal B$, mainly $P_T$ is increased. After the relaxation has run its course, we have $\Delta,P_\Delta=0$,  $P_T=$ const. If one modifies $T_g$ now,  the density will change in response, see Eq.(\ref{2b-5}),   as
\begin{equation}
\frac{\rho^2}{\rho_{cp}-\rho}=\frac{2P}{ab_0T_g^2},
\end{equation}
if $b_1$ is neglected. This happens reversibly, in both directions.  

To calculate the dynamics of irreversible compaction, we start with the relation 
\begin{equation}
\left.\frac{\partial\rho}{\partial\Delta}\right|_{T_g,P}=\frac{\partial_t\rho}{\partial_t\Delta}=\frac{\partial P/\partial\Delta}{\partial P/\partial\rho} \equiv A.
\end{equation}
Because $\partial_t\Delta+\lambda_1T_g\Delta=-v_{kk}=\partial_t\rho/\rho$, see Eq.(\ref{eqD2}), the relaxation of $\Delta$ is given by 
\begin{equation}\label{DRelax}
\partial_t\Delta=-\Delta/\tau_\Delta,\quad \tau_\Delta\equiv(1+A/\rho)/\lambda_1T_g.
\end{equation}
Note $\rho=\rho(\Delta)$ and $\tau_\Delta=\tau_\Delta(\Delta)$ for $P, T_g=$ const, hence the dynamics is not exponential. (We assume vanishing shear strain $v_s,u_s\equiv0$.) 


\subsection{Memory Effects versus Hidden Variables\label{hdvhv}}

Changing $T_g$ midway, at constant $P=P_\Delta+P_T$, with $\Delta$ still finite,
will mainly lead to a change in $\Delta$ (as $\rho$ responds more slowly). 
This disrupts the relaxation of $\Delta$, in essence resetting its initial
condition. This was observed in~\cite{mem} and interpreted as a
memory effect. 
Such ``memory" is
usually a result of hidden variables: When the system
behaves differently in two cases, although all state
variables appear to be the same, we speak of memory or
history-dependence. But an overlooked variable that
has different values for the two cases will naturally
explain the difference. In the present case, the manifest and hidden
variables are  $\rho$ and $\Delta$, respectively.

\subsection{Critical State Under Perturbations\label{A2}}
If we want to render the above considerations quantitative, we need to 
relate the amplitude of  external perturbations to the temperature $T_e$. One clean way to do this is to observe the critical state exposed to external perturbations. 
Inserting Eq.(\ref{TE}) into (\ref{allProp}), we find
\begin{equation}\label{cs2}
\pi_s(\Delta_c,u_c)=\frac{\pi_c}{1+(T_e/fv_s)^2},
\end{equation}
instead of Eqs.(\ref{cs}). Clearly, the critical shear stress vanishes at slow rates, $T_e\gg fv_s$, and is unchanged at higher ones, $T_e\ll fv_s$. This strongly rate-dependent behavior has variably been observed, see eg.~\cite{vHecke2011}.  
Especially, we have $\pi_s=\frac12\pi_c$ for $T_e=fv_s$. Therefore, varying the amplitude of an external perturbation at a given shear rate, we may identify the amplitude at which  $\pi_s=\frac12\pi_c$ holds as equivalent to $fv_s$. 

\subsection{Tapping\label{tapping}}
Next, we consider tapping in greater detail, aiming to relate it to the above presented,  generally valid mechanism of compaction.
Gentle tapping leads to both $\Delta$ and  $T_g$ fluctuating. As long as grains loosen but do not loose contact with one another, $\Delta$ remains finite at all time, $P_T$ may be neglected, and $\bar\Delta$ (averaged in time) will relax monotonically, as accounted for by Eq.(\ref{DRelax}). 
Because a given layer in a granular column with a free upper surface is subject to a constant pressure, the density increases to  compensate for the diminishing $\Delta$. 

Stronger tapping leads to a higher $T_g$, with (1) grains losing contact and $\Delta$ vanishing periodically on one end, and (2) $T_g$ vanishing with $\Delta$ maximal on the other, when grains come down to a crushing stop. This is a hard case to account for, because the system undergoes the transition from solid to liquid, further on to gas and collisionless free flight, then all the way back again, after every tap, see \cite{lupou}. 
And it raises the question whether the system, when being tapped again, will pick up the relaxation of $\Delta$ where it was left at, when the system last crushed to a stop. 
Now, given the fact that tapping is but one way to achieve compaction, leading to results very similar to that of many other methods~\cite{1nico}, it does appear that grains in free flight remember the average packing efficiency when they were at rest. Possibly, this memory is the basic effect, while a finite $T_g$ and a further reduction of $\Delta$ a small perturbation, which  becomes evident only after the accumulation of many taps. 

This is only a surmise, but if proven true, we may take tapping as driven by the same compaction mechanism as described above. Then GSH provides an alternative understanding for compaction  that is transparent, conventional and demystified.

\section{Constant Rate Experiments\label{A}}
Driving a granular system at a given uniform shear rate, with $T_g$ quickly settling into the stationary solution, $T_g=fv_S$, the two relevant relaxation equations are given by Eqs.(\ref{sta1}). 

\subsection{The Hypoplastic Model\label{A-4}}
The hypoplastic model is a state-of-the-art constitutive relation that achieves considerable realism in the rate-independent regime of elasto-plastic motion, especially in triaxial experiments. For $v_{kk}=0$, and since $\sigma_{ij}=\pi_{ij}$, it has the form
\begin{equation}\label{3b-1}
\partial_t\pi_{k\ell}=H_{k\ell ij}v^*_{ij}+
\Lambda_{k\ell}v_s, 
\end{equation} 
as postulated by Kolymbas~\cite{kolymbas1}, where
$H_{k\ell ij},\Lambda_{k\ell}$ are functions of the stress and density. 
Clearly, it has very similar dependence on the shear rate as Eqs.(\ref{sta1}). In fact, taking 
\begin{align*}
\partial_t\pi_{k\ell}(\Delta,u_s)=\frac{\partial\pi_{k\ell}}{\partial u_{ij}}\partial_t u_{ij}=-\frac{\partial^2w_\Delta}{\partial u_{ij}\partial u_{k\ell}} (\partial_tu^*_{ij}-\frac{\delta_{ij}}3\partial_t\Delta),
\end{align*}
with Eqs.(\ref{2b-2},\ref{sta1}), we can easily deduce the expressions for $H_{ijk\ell}$ and $\Lambda_{ij}$. (The equivalence between GSH and hypoplasticity also holds for $v_{kk}, \partial_t\rho\not=0$.)

The name of ``hypoplasticity" arose originally because it was believed that $H_{ijk\ell}$ and $\Lambda_{ij}$ cannot be obtained from any potential (contrary to what  is done here). As a result, hypoplasticity was a very flexible theory, containing $42$ functions as adjustable parameters.  Great efforts have been, and are being, invested in finding accurate expressions for them.
The results of GSH have started to change this among modern practitioners of hypoplasticity. Recently, Niemunis, Grandas Tavera and Wichtmann use a Gibbs potential to obtain $H_{ijkl}$ -- an improvement that they call {\em neo-hypoplasticity}. Calibrating  $H_{ijkl}$ using incremental stress-strain relations, they found excellent agreement with observation of elasto-plastic motion~\cite{99}.  Subsequently, we showed that this Gibbs potential is, legendre transformed, rather similar to $w_\Delta$ of Eq.(\ref{2b-2}), see~\cite{ActaGeoMechanica}.
This is reassuring, as it implies that all results of hypoplasticity -- including various butterfly-curves, the approach to the critical state and different slopes at load and unload -- can also be produced by GSH. 

\subsection{Load and Unload\label{A-1}}
One big advantage of GSH is its simple structure that enables analytic solutions for many experimental situations. Therefore, in spite of the general agreement with hypoplasticity, we shall examine Eqs.(\ref{sta1}) more closely.
For load and unload,  the equations are,  respectively,
\begin{equation}
\partial_tu_s=v_s-\lambda fu_sv_s,\quad \partial_tu_s=-v_s-\lambda fu_sv_s.
\end{equation}
Note, first of all, that the described behavior is elastic for vanishing shear stress, $u_s\to0$. The shear strain is maximal in the stationary,  critical state, in which both terms cancel, and dissipation, produced by the relaxing second term, is considerable. (The shear strain may well be non-monotonic, passing a maximum before assuming the critical value, see next section.) 
Next, note that the slope $u_s$ versus $v_s$ is, respectively for load and unload, $1-\lambda fu_s$ and $-(1+\lambda fu_s)$, a simple fact not related to any ``history-dependence."

\subsection{Approach to the Critical State\label{A-2}}
Solving Eqs~(\ref{sta1}) for constant 
$v_s$, with the initial conditions: $\Delta=\Delta_0, u_s=0$, the relaxation into the critical state is an exponential decay for $u_s$, and a sum of two for $\Delta$,
\begin{align}\label{3b-6}
u_s(t)=u_c(1-e^{-\lambda f\varepsilon_s}),\quad \varepsilon_s\equiv v_st,
\\\nonumber \Delta(t)=\Delta_c(1+f_1\,e^{-\lambda f
\varepsilon_s}+f_2e^{-\lambda_1f\varepsilon_s}), \\\nonumber
f_1\equiv\frac{\lambda_1}{\lambda-\lambda_1},\quad
f_2\equiv\frac{\Delta_0}{\Delta_c}-\frac{\lambda}{\lambda-\lambda_1}.\end{align}
It is useful that a simple, analytical solution for this signature experiment of soil mechanics exists. Typically, $\lambda$ is larger than $\lambda_1$, and the decay of $u_s$ and $f_1$ faster than that of $f_2$. Note $f_2$ may be
negative, and $\Delta(t)$ is then non-monotonic. The associated pressure  $P_\Delta$ 
and shear stress $\pi_s$ are those of Eqs~(\ref{2b-2b}), neither is monotonic for a negative $f_2$. 

\subsection{Shear Jamming\label{A-3}}
Generally speaking, that grains jam when the volume is reduced seems obvious, less so when they are sheared. Yet this has been observed~\cite{SJ1}. Further, the density at which this happens seems to memorize the deformation history. After a cyclic compression, the density at which the grains unjam is significantly larger~\cite{SJ2,SJ3}.

In GSH, a state is  jammed if it stably sustains an elastic stress,  $\Delta,u_s\not=0$. Between the random lose and random close values, $\rho_{lp}\le\rho\le\rho_{cp}$, the system is unjammed if $\Delta,u_s$ vanish -- independent of the density.  Note that for $\rho<\rho_{lp}$, we always have  $\Delta,u_s\equiv0$, and the system is never jammed; while for $\rho>\rho_{cp}$,  $\Delta$ is always finite, and the system cannot be unjammed.

Shear jamming is a special case of the approach to the critical state, as accounted for by  Eqs.(\ref{3b-6}).  
Starting from $\Delta_0=0$ (rather than  $\Delta_0\not=0$ as usual),  a steady shear rate $v_s$ will (for $\rho_{lp}\le\rho\le\rho_{cp}$) increase $\Delta, u_s$, jam the system, and push it eventually into the critical state. This is the simple physics of shear-jamming. 

The memorized jamming density is easily explained by  granular compaction in the presence of $T_g$, as considered in appendix B.  
A cyclic compression increases the packing efficiency, because it 
produces a $T_g$ that lets  $\Delta, u_s$ relax. When the system 
becomes unjammed again, with $\Delta,u_s=0$, the actual density is larger than the initial one. 
Calling a density at which $\Delta,u_s$ become zero  the {\it jamming density}, we find it to possess memory. As in appendix B.2, this is due to the hiden variables,  $\Delta,u_s$. 

Finally, a word on the intermediate so-called {\it fragile state} that is sandwiched between the unjammed and  stably jammed one.  Starting from $\Delta_0=0$ and $u_s=0$, the growth of $u_s$ is initially faster than that of $\Delta$, see Eqs.(\ref{sta1}), such that the stability condition $u_s/\Delta\le\sqrt{2{\cal B}/{\cal A}}$ of Eqs.(\ref{2b-3})
 is breached. Only after $u_s$ has reached a certain size, does $\Delta$ catch up, to satisfy  Eq.(\ref{2b-3}) and jam the system stably. 
[In usual geotechnical experiments, the initial condition always has a sufficiently large $\Delta_0$ such that Eqs.(\ref{2b-3}) is  never breached.] GSH at present does not provides an answer for how a granular system behaves when stability conditions are breached, but we are working to include this aspect.

\section{ Constant Stress Experiments\label{B}}
We now consider the case of constant stress and density, accounted for by Eqs.(\ref{sta2}), a relaxation-diffusion  equation for $fv_s=T_g\sqrt{\pi_s/\pi_c}$. 

In this context, it is useful to realize that stress-controlled experiments cannot be performed in simple triaxial apparatuses with stiff steel walls, because the correcting rates employed by the feedback loop to keep the stress constant (or slowly changing) are usually too strong. As a result, $T_g$ is excited that distorts, even overwhelms, its relaxation. The situation is then more one of consecutive constant rates, less one of constant stress. Employing soft springs to couple the granular system to its driving device makes small-amplitude stress corrections possible without exciting much $T_g$.

\subsection{Diverging Shear Strain\label{B-1}}
First, consider uniform solutions of  Eqs.(\ref{sta2}).
With the initial conditions $T_g=T_0$, $v_s=v_0$ and $T_0\sqrt{\pi_s/\pi_c}=fv_0$  at $t=0$, we find 
\begin{equation}\label{3b-10} 
T_g={T_0}/({1+{r_T}{T_0}t}),\quad r_T\equiv R_T[1-\pi_s/\pi_c]. 
\end{equation}
The solution for the shear rate, 
$v_s={v_0}/(1+ r_vv_0 t)$, with $r_v\equiv {r_T} f\sqrt{\pi_c/\pi_s}$,
implies a logarithmically divergent total shear strain,
\begin{equation}
\varepsilon_s-\varepsilon_0\equiv\int v_s{\rm d} t=\ln(1+r_vv_0t)/r_v.
\end{equation}
Note that $\varepsilon_s$ does not actually diverge, 
because as $T_g\to0$, the system enters the quasi-elastic regime, 
where its relaxation becomes exponential,  see~\cite{granR4}. So even if the creep $\varepsilon_s$ is large close to $\pi_c$, it comes to a halt eventually, and the system is mechanically stable. 

The $T_g$-relaxation is slower the closer  $\pi_s$ is to $\pi_c$, infinitely so  for  $\pi_s=\pi_c$. 
The state is then indistinguishable from the rate-controlled, critical state, which, clearly,  may be  maintained also at given stress. For $\pi_s>\pi_c$, the relaxation rate is negative, and $T_g\propto v_s$ will  grow without bound. The system then has two possibilities to become stable again, either a non-uniform shear band, or accelerating to the rate-dependent regime of the $\mu(I)$-rheology, see the sections below. 

In an experiment involving a fan submerged in sand and coupled to the motor with a very soft spring, Nguyen et al.~\cite{aging} first pushed the system 
to a certain shear stress at a given rate, producing a $T_g$. Then, switching to maintaining the shear stress, they observed a shear strain  $\varepsilon_s(t)$ that appears to diverge logarithmically. 

Ignoring the strongly non-uniform stress distribution,  they carried out  an analysis using two scalar equations 
borrowed from glassy media that may be roughly mapped to the ones employed above, where the granular temperature $T_g$,  its relaxation rate $R_T$, and its production rate 
$R_Tf^2$ were replaced by {\em fluidity, aging} and {\em rejuvenation parameter}. 
This is certainly a sensible approach when one lacks a granular theory to interpret the data. But less so given the fact that GSH provides a unified, tensorial and realistic treatment tailored to granular media, thus affording a transparent, well-founded and innate understanding. One circumstantial support for GSH are {hot spots} --  localized regions of strong, temporary deformations that the authors observed. They identified the rate of hot spots as fluidity, but could more directly have taken them as  $T_g$.

Finally, we note that an infinitesimal $T_g$ will not destabilize a static shear stress exceeding $\sigma_c$, only a $T_g$ sufficiently large will grow without bound. This is because the critical stress diverges for $T_g\to0$ and the window between $\sigma_c$ and $\sigma_s^{yield}$ vanishes~\cite{granR4,granRexp}. Therefore, a static shear stress remains metastable for $T_g=0$, turning instable only at  the yield stress, $\pi_s^{y}/\sqrt2 P=\sqrt{{\cal A}/{\cal B}}$, as given in Eq.(\ref{2b-3}).  

\subsection{Angle of Repose $\varphi_{re}$ versus Angle of Stabilty $\varphi_{st}$ \label{B-2}}
Aranson and Tsimring were the first to construct a theory for these two angles~\cite{aranson,Aranson1}. Taking the stress as the sum of a solid- and a  fluid-like part, they define an 
order parameter  $\hat\varrho$ that is 1 for solid, and 0 for dense flow. They then postulate a free energy $f(\hat\varrho)$ such that it is stable with $\hat\varrho=1$ only for $\varphi< \varphi_{st}$, with $\hat\varrho=0$ only for $\varphi>\varphi_{re}$, and  $\varphi_{st}>\varphi>\varphi_{re}$ as the bi-stable region. 
In GSH, these two angles need not be postulated {\it ad hoc}, as they  are already given by the yield and critical stress, from Eqs.(\ref{2b-3},\ref{cs}): 
\begin{align}\label{sb12}
\tan\varphi_{st}&=\pi_{yie}/\sqrt2 P_\Delta=\sqrt{{\cal A}/{\cal B}},
\\
\tan\varphi_{re}&={\sigma_c}/\sqrt2\,P_c, \quad\varphi_{re}<\varphi_{st}.
\end{align}
The collapse that occurs when one slowly tilts a plate supporting a layer of grains at $\varphi_{st}$, is a process that happens at $T_g=0$ and hence encoded in $w_\Delta$.

The angle of repose  $\varphi_{re}$ is related to the results of the last section: As long as the shear stress is held below the critical one, $\pi_s<\pi_c$, the $T_g$-relaxation will run its course, and the system is in a static, mechanically stable state  afterwards. At  $\pi_s=\pi_c$, the system becomes critical, and no longer comes to a standstill. Therefore, $\varphi_{re}$ is given by the critical stress.

The inequality $\varphi_{re}<\varphi_{st}$ holds because the critical state is an elastic solution, cf. Eq.(\ref{cs}), and $\varphi_{st}$ the angle at which all elastic solutions become unstable.

\subsection{Shear Bands\label{B-3}}
For $\pi_s>\pi_c$, no uniform stationary solution of Eq.(\ref{sta2}) exists, but a nonuniform one does, with $T_g\equiv0$  for $x\le0$ and $x\ge\xi_{sb}$, and a shear band  in between:
\begin{align}\nonumber
\nabla^2 T_g&=-T_g/\xi^2_{sb},\quad{\xi}^2_{sb}\equiv\xi_T^2/[\pi_s/\pi_c-1],
\\
 v_s/v_s^0&=T_g/T_g^0=\sin(\pi x/\xi_{sb}). \label{nsb2}
\end{align}			
The velocity difference across the band is $\Delta v=\int v_s{\rm d}x=\int v_s^0\sin(\pi x/\xi_{sb}){\rm d}x$, or
\begin{equation}
v_s^0 =\Delta v/(2\xi_{sb})=\sqrt{\pi_s/\pi_c}\,T_g^0/f.
\end{equation}
That the correlation length $\xi_{sb}$ diverges for $\pi_s=\pi_c$ gives a justification for the term {\it critical.}
Note the density $\rho$ must be lower in the shear band than in the  quiescent region, $\rho_{sb}<\rho_{q}$, because only then can the shear stress $\pi_s$ (in 1D a constant quantity) be smaller in the quiescent region, but larger in the shear band,  than the critical stress, $\pi_c(\rho_{q})>\pi_s>\pi_c(\rho_{sb})$. 

Due to the lacks of a length scale, constitutive models such as hypoplasticicty or barodesy cannot account 
for shear bands. There are various approaches to ameliorate it, by introducing gradient terms~\cite{wu2} or adding state variables to account for the couple stress and the Crosserat rotation~\cite{wu1}.
The latter method works, but at the price of a  far more complex theory, constructed for the sole purpose of 
describing shear bands. And it poses questions about the physics: If the couple stress is important in the shear band, because it is fluid, why is it not in the uniformly fluid and gaseous state of granular media?

\section{Fast Dense Flow}
Eqs.(\ref{sta2}) state that for subcritical stresses: $\sigma_s=\pi_s<\pi_c$, the granular temperature and the shear rate, $T_g=fv_s$, will relax to zero; for super-critical stresses,   $\sigma_s=\pi_s>\pi_c$, they will accelerate, until Eqs.(\ref{27}),  again a stable solution,  are satisfied.  

\subsection{The ${\mu(I)}$-Rheology\label{C}}

Although $\sigma_s$ and $P$ are functions of $\rho,v_s^2$, we can rewrite Eq.(\ref{27}), with 
\begin{equation}
\mu\equiv\frac{\sigma_s}P,\quad \mu_1\equiv\frac{\pi_c(\rho)}{P_c(\rho)},\quad
\mu_2\equiv \frac{e_s(\rho)}{e_p(\rho)}
\end{equation}
as
\begin{equation}\label{gsh-mu}
\mu=\mu_1+(\mu_2-\mu_1)\hat I,\quad \hat I\equiv\frac{e_pv_s^2}P=\frac{P_T}{P_T+P_c},
\end{equation}
where both $\mu_1$ and $\mu_2$ have been observed as independent of the density, see~\cite{Bagnold,schofield,wood1990}. [Note 
$e_p=\frac12f^2\rho^2\partial b/\partial\rho$, see Eqs.(\ref{2b-5-2},\ref{27}).] 

Stating that $\mu$ varies between two density-independent plateaus with a single variable $\hat I$ that vanishes with $v_s$, and goes to one as $v_s\to\infty$, this GSH-formula is close to the well-known $\mu(I)$-rheology~\cite{pouliquen1,pouliquen2,pouliquen3,DEM-mu1,DEM-mu2}:  $\mu=\mu_1+({\mu_2-\mu_1})I/({I + I_0})$, with $\mu_1,\mu_2$ density-independent, and $I/(I + I_0)$ going from zero to one. 

But there is the difference that $\hat I\propto v_s^2/P\propto I^2$. 
Supporting GSH is the  rate-independence of the  critical state,
\[\partial\sigma_s/\partial v_s\stackrel{v_s\to0}\equiv0,\]
valid for many orders of magnitude of $v_s$. If the stress is an analytic function of $v_s$, with no kink, linear dependence $\partial\sigma_s/\partial v_s=$ const. is ruled out, but not a quadratic one. [Dimensional arguments lead to both the  $\mu(I)$-rheology and to Eq.(\ref{gsh-mu}).]

For  $\rho<\rho_{\ell p}$, there is no elastic solution, $P_c,\pi_c=0$. Then $\mu\equiv\mu_2$ irrespective how large the shear rate $v_s$ is. 
When studying granular rheology, one varies $ v_s$ keeping  either the density or the pressure constant. Since $P_c\gg e_p(\rho) v_s^2$ and $\pi_c\gg e_s(\rho) v_s^2$ for the usual rates, it is not easy to go beyond the limit of $\mu=\mu_1$ for given $\rho$. Not so for given pressure, because the density decreases for increasing $v_s$, and a discontinuous transition from  Eqs.(\ref{gsh-mu}) to $\mu\equiv\mu_2$ takes place at $\rho=\rho_{\ell p}$, where $P, \sigma_s$ decrease by three orders of magnitude, see~\cite{campbell}. 
The $\mu(I)$-relation, lacking this bit of information, is in fact valid only for  $\rho>\rho_{\ell p}$. 

On the other hand, the second part of  $\mu(I)$-rheology, concerning 
the packing fraction $\phi\equiv \rho/\rho_g$ and taken as $\phi=\phi(I)$, holds only for  $\rho<\rho_{\ell p}$, because $\phi=f(I)$, or $f^{-1}(\phi)=I\propto v_s^2/P$, implies $P_c=0$. 

For given density $\rho$, the shear stress $\sigma_s=\mu_1P^c(\rho)+\mu_2e_p(\rho) v_s^2$ is a monotonic function of the rate~$v_s$. For constant pressure, 
with $\rho=\rho(P, v_s)$ given by $P=P^c(\rho)+e_p(\rho) v_s^2$, the shear stress $\sigma_s$  may well be non-monotonic~\cite{exp6.}.

\subsection{The Solid-Liquid Boundary\label{TgRelaxation}}
One frequently observes granular liquid executing fast dense flow, bordering on a quiescent, or solid, region. More careful observation reveals that the velocity is continuous, with a quickly decaying creep taking place in the solid, see Komatsu et al~\cite{komatsu}, Crassous et al~\cite{crassous}. GSH accounts for this as a result of $T_g$ diffusing from the fluid region into the solid one. 

We take the solid-liquid boundary at $x=0$, and only variations perpendicular to it. For such a one-dimensional geometry, the pressure $P$ and shear stress $\sigma_s$ are necessarily uniform. The density  is discontinuous at $x=0$, being larger on the solid side, $\rho_s>\rho_\ell$. Hence $\sigma_s>\pi_c$ in liquid, executing fast dense flow according to Eq.(\ref{gsh-mu}), and  $\sigma_s=\pi_s<\pi_c$ in solid.

The shear rate $T_g=fv_s$ is uniform on the liquid side, for $x<0$. On the solid side, $x>0$, with the liquid maintaining the shear stress $\sigma_s$, Eqs.(\ref{sta2}) hold.  Therefore
\begin{align}
\label{cr1}
\nabla^2T_g=T_g/\xi^2_{cr},\quad \xi_{cr}^2\equiv\xi_T^2/[1-\pi_s/\pi_c]
\\\label{cr2}
\text{implying}\quad v_s/v_s^{0}= T_g/T_g^{0}= \exp(-x/\xi_{cr}),
\end{align}
where $T_g^{0}$ is the fluid value, and $v_s^{0}=[T_g^{0}/f(\rho_s)]\sqrt{\pi_s/\pi_c(\rho_s)}$. (The velocity and $T_g$ are continuous, the rate $v_s$ is not.) We require $T_g\to0$ for $x\to\infty$.

It is not surprising that the decay length $\xi_{cr}$  diverges for $\pi_s=\pi_c$, because the solid region turns critical then, and ceases to exist. 

\subsection{Kamrin's nonlocal constitutive relation\label{B-4}}

In two recent papers~\cite{kamrin,kamrin2}, Kamrin et al propose a nonlocal constitutive relation (KCR) well capable of accounting for the steady flows in the split-bottom cell~\cite{fenistein}. A key ingredient is the fluidity $g\equiv v_s/\mu$. Denoting $\mu\equiv\sigma_s/P$, $\mu_s\equiv\sigma_c/P_c$, it is taken to obey
\begin{equation}
\xi^2_{cr}\nabla^2g=g-g_{loc},\quad \xi_{cr}\propto1/\sqrt{|\mu-\mu_s|}.
\end{equation}
Because $g_{loc}=0$ for $\mu<\mu_s$, this relations is rather similar to Eq.(\ref{cr1}), with $g$ assuming the role of $T_g$, and the two decay lengths diverging at the same stress value.
For $\mu\geq\mu_s$, the system is fluid and $g=g_{loc}$ essentially constant. With $g_{loc}\propto\sqrt P(1-\mu_s/\mu)$, KCR is consistent with a first-order expansion in the inertial number of the $\mu(I)$-rheology. 
Therefore, GSH is, in this limit, also similar to KCR.  

This is fortunate,  because again, there is a symbiotic relation between GSH and KCR, similar to that between GSH and hypoplasticity, which gives the constitutive relations a solid foundation in physics, and GSH a robust connection to reality.

More recently, Kamrin and Bouchbinder constructed a ``two-temperature continuous mechanics,'' with  $\theta_c$ and $\theta_v$, the first a configuration temperature, the second for the vibrational degrees of freedom~\cite{ttcm}.  (Microscopic degrees of freedom are not considered. There would be three temperatures otherwise.)  They hypothesize that the fluidity  $g$ may be related to   $\theta_c$. We believe that $g$ is, as discussed above, essentially $T_g$ of GSH, which (quantifying the fluctuating granular degrees of freedom) is closer to  $\theta_v$. 


\begin{thebibliography}{99}
\bibitem{CundallStrack}P.A. Cundall, and O.D.L. Strack,  “A discrete numerical model for granular assemblies”,
{\it Geotechnique}, {\bf 29} No. 1, pp. 47-65 (1979).

\bibitem{dem1}H. J. Herrmann and S. Luding.
Review Article: Modeling granular media with the computer, 
Continuum Mechanics and Thermodynamics 10 (4) 189-231 (1998)

\bibitem{dem2}J.N. Roux
AIP Conf. Proc. 1542, 46 (2013), 
{\it``Quasistatic behaviour of granular materials: Some things we learned from DEM studies,''} {http://dx.doi.org/10.1063/1.4811865}


\bibitem{kin0}
P.~K. Haff.
\newblock {\it Journal of Fluid Mechanics Digital Archive}, 134(-1):401--430,
  1983.
\bibitem{kin1} J. T. Jenkins and S. B. Savage, {\it J. Fluid Mech.} {\bf 130}, 187 (1983).
\bibitem{kin2} 
S. B. Savage, {\it Adv. Appl. Mech.} {\bf 24}, 289 (1984).
\bibitem{kin3} 
C.S. Campbell, {\it Ann. Rev. Fluid Mech.} {\bf 22}, 57 (1990).
\bibitem{kin4} I. Goldhirsch, {\it Chaos} {\bf 9}, 659 (1999) and  {\it Annu. Rev. Fluid Mech.} {\bf 35}, 267 (2003).




\bibitem{Edw} S.F. Edwards, R.B.S. Oakeshott,
Physica A{\bf 157}, 1080 (1989); 
\bibitem{raphi}R. Blumenfeld, J.F. Jordan, S.F. Edwards. Phys. Rev. Lett.{\bf 109}, 238001 (2012)

\bibitem{1nico} P. Richard, M. Nicodemi, R. Delannay,
P. Ribiere, D. Bideau, Nature, {\bf 4}, 121 (2005)

\bibitem{1}
A. Baldassarri, A. Barrat, G. D’Anna, V. Loreto, P. Mayor and
A. Puglisi. J. Phys.: Condens. Matter 17 (2005) S2405–S2428  


\bibitem{2}
Dapeng Bi, Silke Henkes, Karen E. Daniels, Bulbul Chakraborty. The Statistical Physics of Athermal Materials. {\it Annual Review of Condensed Matter Physics} 6: 63-83 (2015); or 
arXiv:1404.1854, 2014




\bibitem{granR1}Y.M.~Jiang and M.~Liu. Eur. A brief review of granular elasticity. Phys. J. E 22, 255 (2007).

\bibitem{granR2}
Y.M.~Jiang and M.~Liu.
\newblock Granular solid hydrodynamics.
\newblock {\em Granular Matter}, 11:139, May 2009.
Free download: {\scriptsize
www.springerlink.com/content/a8016874j8868u8r/fulltext} 

\bibitem{granR3}
Y.M.~Jiang and M.~Liu.
\newblock The physics of granular mechanics.
\newblock In D.~Kolymbas and G.~Viggiani, editors, {\em Mechanics of Natural
  Solids}, pages 27--46. Springer, 2009.

\bibitem{granR4}  
  Y.M.~Jiang and M.~Liu.
  \newblock Granular Solid Hydrodynamics  (GSH): a broad-ranged macroscopic theory of granular media.
\newblock {\em Acta Mech.} 225, 2363–2384 (2014)
DOI 10.1007/s00707-014-1131-3.

\bibitem{granRgudehus}
G.~Gudehus, Y.M. Jiang, and M.~Liu.
\newblock Seismo- and thermodynnamics of granular solids.
\newblock {\em Granular Matter}, 1304:319--340, 2011.

\bibitem{granL1}Y.M.~Jiang, M.~Liu. {\it Phys. Rev Lett,} {\bf 91}, 144301(2003).
\bibitem{granL2}Y.M.~Jiang, M.~Liu. {\it Phys. Rev Lett,} {\bf 93}, 148001(2004).
\bibitem{granL3}Y.M.~Jiang, M.~Liu. {\it Phys. Rev Lett,} {\bf 99}, 105501 (2007).
\bibitem{exp1.}	Krimer, D.O., Pfitzner, M., Bräuer, K., Jiang, Y., Liu, M.: Granular elasticity: general considerations and the stress dip in sand piles. Phys. Rev. E 74, 061310 (2006)
\bibitem{exp2.}	Bräuer, K., Pfitzner, M., Krimer, D.O., Mayer, M., Jiang, Y., Liu, M.: Granular elasticity: stress distributions in silos and under point loads. Phys. Rev. E 74, 061311 (2006)
\bibitem{exp3.}	Jiang, Y.M., Liu, M.: Incremental stress-strain relation from granular elasticity: Comparison to experiments. Phys. Rev. E 77, 021306 (2008)
\bibitem{exp4.}	Y.M. Jiang and M. Liu. GSH, or Granular Solid Hydrodynamics: on the Analogy
between Sand and Polymers. AIP Conf. Proc. 7/1/2009, Vol. 1145 Issue 1, p1096.
\bibitem{exp5.}	Mayer, M., Liu, M.: Propagation of elastic waves in granular solid hydrodynamics. Phys. Rev. E 82, 042301 (2010)
\bibitem{exp6.}	D. Krimer, S. Mahle and M. Liu, Dip of the Granular Shear Stress, Phys. Rev. E86, 061312 (2012)
\bibitem{exp7.}	Jiang, Y.M., Zheng, H.P., Peng, Z., Fu, L.P., Song, S.X., Sun, Q.C., Mayer, M., Liu, M.: Expression for the granular elastic energy. Phys. Rev. E 85, 051304 (2012)
\bibitem{exp8.}	Zhang, Q., Li, Y.C., Hou, M.Y., Jiang,Y.M., Liu, M.: Elastic waves in the presenceof a granularshear band formed by direct shear. Phys. Rev. E 85, 031306 (2012).
\bibitem{exp9.}	Jiang, Y., Liu, M.: Proportional path, barodesy, and granular solid hydrodynamics. Granul. Matter 15, 237–249 (2013)
\bibitem{PG2013}
Yimin Jiang and Mario Liu, 
AIP Conf. Proc. 1542, 52 (2013); {\it Stress- and rate-controlled granular rheology,} http://dx.doi.org/10.1063/1.4811867
\bibitem{granRexp}
  Y.M.~Jiang and M.~Liu. {\it Applying GSH to a wide range of experiments
in granular media.} Eur. Phys. J. E (2015) 38:15

\bibitem{kolymbas1}
D.~Kolymbas.
\newblock {\em Introduction to Hypoplasticity}.
\newblock Balkema, Rotterdam, 2000.
\bibitem{kolymbas2} W.~Wu and D.~Kolymbas.
\newblock {\em Constitutive Modelling of Granular Materials}.
\newblock Springer, Berlin, 2000.


\bibitem{kamrin}D.L. Henann and K. Kamrin. {\it Proceedings of  the National  Academy of  Sciences},{\bf 110}, 6730 (2012). http://www.pnas.org/content/110/17/6730.full. 
\bibitem{kamrin2} K. Kamrin and G. Koval. {\it Phys.Rev.Lett.} {\bf108},178301 (2012)

\bibitem{pouliquen1} GDR MiDi, {\it Eur. Phys. J.} {bf E 14}, 341, (2004).
\bibitem{pouliquen2}
Yo\"{e}l Forterre and Olivier Pouliquen.
\newblock Flows of dense granular media.
\newblock {\em Annu. Rev. Fluid Mech.}, 40:1--24, 2008.

\bibitem{pouliquen3} 
Y. Forterre and O. Pouliquen, {\it Annu. Rev. Fluid Mech.} {\bf 40}, 1,
(2008).


\bibitem{SJ1}D.P. Bi, J. Chang, B. Chakraborty, R.P. Behringer, {\it Nature}, {\bf 480}, 355 (2011). Somayeh Farhadi and Robert P. Behringer. Dynamics of sheared ellipses and circular disks: effects of particle shape. Phys. Rev. Lett., 112:148301, 2014. 

\bibitem{SJ2}N. Kumar, Stefan Luding, Memory of jamming -- multiscale models for soft and granular matter,
{\it Granular Matter} 18, 58, 2016 
\bibitem{SJ3}S. Luding, Granular matter: So much for the jamming point,
{\it Nature Physics} 12, 531-532, 2016
\bibitem{aging}
Van~Bau Nguyen, Thierry Darnige, Ary Bruand, and Eric Clement.
\newblock Creep and fluidity of a real granular packing near jamming.
\newblock {\em Phys. Rev. Lett}, 107:138303, 2011.


\bibitem{Khal}
I.~M. Khalatnikov.
\newblock {\em Introduction to the Theory of Superfluidity}.
\newblock Benjamin, New York, 1965.
\bibitem{LL6}
L.~D. Landau and E.~M. Lifshitz.
\newblock {\em Fluid Mechanics}.
\newblock Butterworth-Heinemann, 1987.
\bibitem{deGennes}
P.G. de~Gennes and J.~Prost.
\newblock {\em The Physics of Liquid Crystals}.
\newblock Clarendon Press, Oxford, 1993.


\bibitem{kubo}R. Kubo, Rep. Prog. Phys. {\bf 29} 255 (1966)


\bibitem{Wichtmann09}
T.~Wichtmann, A.~Niemunis, and T.~Triantafyllidis.
\newblock On the deter-
mination of a set of material constants for a high-cycle accumulation
model for non-cohesive soils.
\newblock {\em Int. J. Numer. Anal. Meth. Geomech.}, 
2010; 34:409–440


\bibitem{Houlsby2}
I.~F. Collins and G.~T. Houlsby.
\newblock Application of thermomechanical principles to the modelling of
  geotechnical materials.
\newblock {\em Proc. R. Soc. Lond. A}, 453:1975--2001, 1997.


\bibitem{polymer-1} H. Temmen, H. Pleiner, M. Liu and H.R. Brand, {\it
Convective Nonlinearity in Non-Newtonian Fluids,} Phys. Rev. Lett. {\bf
84}, 3228 (2000). 
\bibitem{polymer-2}H. Temmen, H. Pleiner, M. Liu and H.R.
Brand,{\it Temmen et al. reply}, Phys. Rev. Lett. {\bf 86}, 745 (2001).
\bibitem{polymer-3}
Oliver M\"uller, Mario Liu, Harald Pleiner, and Helmut R. Brand
{\it Transient elasticity and polymeric ﬂuids: Small-amplitude deformations}
Phys.Rev. {\bf E 93}, 023113 (2016)
\bibitem{polymer-4}Oliver M\"uller, Mario Liu, Harald Pleiner, and Helmut R. Brand
{\it Transient elasticity and the rheology of polymeric ﬂuids with large amplitude deformations}
Phys.Rev. {\bf E 93}, 023114 (2016)



\bibitem{hardin}
B.O. Hardin and F.E. Richart.
\newblock Elastic wave velocities in granular soils.
\newblock {\em J. Soil Mech. Found. Div. ASCE}, 89: SM1:33--65, 1963.
\bibitem{luding2009}
Stefan Luding.
\newblock Towards dense, realistic granular media in 2d.
\newblock {\em Nonlinearity}, 22:101--146, 2009.
\bibitem{Bocquet}  
L.~Bocquet, W.~Losert, D.~Schalk, T.~C. Lubensky, and J.~P. Gollub.
\newblock Granular shear flow dynamics and forces: Experiment and continuum
  theory.
\newblock {\em Phys. Rev. E}, 65(1):011307, Dec 2001.

\bibitem{vHecke2011}
J.A. Dijksman, G.H. Wortel, L.T.H. van Dellen,
O. Dauchot, and M. van Hecke. 
\newblock Jamming, yielding, and rheology of weakly vibrated granular media.
\newblock {\em Phys. Rev. Lett.}, {\bf 107},
108303(2011).

\bibitem{mem}C. Josserand, A.V. Tkachenko, D.M.
Mueth, H.M. Jaeger, Phys. Rev. Lett., \textbf{85}, 3632
(2000)
\bibitem{lupou} S. Luding, M. Nicolas, O. Pouliquen,
A minimal model for slow dynamics: Compaction of granular media under vibration or shear
page 241 in: Compaction of Soils, Granulates and Powders, D. Kolymbas and W. Fellin (eds.), Balkema, Rotterdam (2000),



\bibitem{99} Andrzej Niemunis, Carlos E. Grandas Tavera, and Torsten Wichtmann:
\newblock{\em Peak stress obliquity in drained and undrained
sands. Simulations with neohypoplasticity,} T. Triantafyllidis (ed.), Holistic Simulation of Geotechnical
Installation Processes, Lecture Notes in Applied and Computational
Mechanics 80, DOI 10.1007/978-3-319-23159-4-5, Springer (2016)
\bibitem{ActaGeoMechanica}  Jiang and Liu: {\it Similarities between GSH, hypo plasticity and KCR.} 
Acta Geotech (2016) 11:519-537, DOI 10.1007/s11440-016-0461-9.


\bibitem{aranson}
I.~S. Aranson and L.~S. Tsimring.
\newblock {\em Phys. Rev. E}, 65:061303, 2002.
\bibitem{Aranson1}
I.~S. Aranson and L.~S. Tsimring.
\newblock {\em Rev. Mod. Phys.}, 78:641, 2006.
\bibitem{wu2}Wei Wu. {\em On high-order hypoplastic models for granular materials}. Journal of Engineering Mathematics {\bf 56}: 23–34 (2006) 

\bibitem{wu1}Tejchman, J. and Wu, W. {\it FE-investigations of micro-polar boundary conditions along interface between soil and structure}, Granular Matter, {\bf 12}, 399 (2010)

\bibitem{Bagnold}
R.~A. Bagnold.
\newblock Experiments on a gravity-free dispersion of large solid spheres in a
  {N}ewtonian fluid under shear.
\newblock {\em Proceedings of the Royal Society of London. Series A.
  Mathematical and Physical Sciences}, 225(1160):49--63, 1954.

\bibitem{schofield}
P.~Wroth A.~Schofield.
\newblock {\em Critical State Soil Mechanics}.
\newblock McGraw-Hill, London, 1968.


\bibitem{wood1990}
D.~M. Wood.
\newblock {\em Soil Behaviour and Critical State Soil Mechanics}.
\newblock Cambridge University Press, 1990.

\bibitem{campbell} C. S. Campbell, {\it J. Fluid Mech.} {\bf 465}, 261 (2002).


\bibitem{DEM-mu1}A. Singh, K. Saitoh, V. Magnanimo, and S. Luding,
Role of gravity or confining pressure and contact stiffness in granular rheology,
{\it New J. Phys.} 17, 043028, 2015

\bibitem{DEM-mu2}S. Roy, S. Luding, and T. Weinhart,
Towards a general(ized) shear thickening rheology of wet granular materials under small pressure, submitted to NJP, 2016

\bibitem{komatsu}
T.S. Komatsu, S.~Inagaki, N.~Nakagawa, and S.~Nasuno.
\newblock Creep motion in a granular pile exhibiting steady surface flow.
\newblock {\em Phys. Rev. Lett.}, 86:1757 1760, 2001.

\bibitem{crassous}
J~Crassous, J-F Metayer, P~Richard, and C.~Laroche.
\newblock Experimental study of a creeping granular flow at very low velocity.
\newblock {\em J. Stat. Mech.}, 2008:P03009, 2008.

\bibitem{fenistein} D. Fenistein, J.W. van de Meent, M.van Hecke, {\it Nature}, {\bf 425} 695 (2003); {\it Phys.Rev.Lett.} {\bf 96}, 118001 (2006); {\bf 96}, 038001 (2006). 

\bibitem{ttcm}Ken Kamrin and Eran Bouchbinder.   {\it Journal of the Mechanics and Physics of Solids,} {\bf 73} 269–288  (2014). {\it Two-temperature continuum thermomechanics of deforming amorphous solids.}
\end{thebibliography}
\end{document}